\documentclass[article]{jfm}

\usepackage[utf8]{inputenc}
\usepackage[T1]{fontenc}
\usepackage{graphicx}
\usepackage{subfigure}
\usepackage{amsmath,bm}
\usepackage{amsfonts}
\usepackage{color}
\usepackage[breaklinks,colorlinks = true,linkcolor = blue,urlcolor=blue,citecolor=blue]{hyperref}
\usepackage[capitalise]{cleveref}
\usepackage{graphicx}
\usepackage{xcolor}
\usepackage{newtxtext}
\usepackage{newtxmath}
\usepackage{natbib}

\newcommand{\C}{\mathbf{C}}
\def \trC  {\text{Tr}[\C]}

\newcommand{\calp}{\mathcal{P}}
\newcommand{\bu}{\mathbf{u}}
\newcommand{\ri}[1]{{\rm #1}}

\newcommand{\ub}{{\bm u}}
\newcommand{\bx}{{\bm x}}
\newcommand{\br}{{\bm r}}
\newcommand{\uh}{\hat{u}}
\newcommand{\uk}{u_k}
\def \ck {C_k}

\newcommand{\bk}{{\bm k}}

\newcommand{\ch}{\hat{C}}

\newcommand{\I}{\mathbf{I}}
\newcommand{\rp}{{\rm p}}
\newcommand{\rf}{{\rm f}}
\def \caldp {\mathcal{D}_\rp}
\def \caldf {\mathcal{D}_\rf}
\newcommand{\kurt}{\mathcal {K}}
\newcommand{\pt}{\partial_t}

\def \pari {\partial_i}
\def \parj {\partial_j}
\newcommand{\nab}{\nabla}

\newcommand{\nabs}{\nabla^2}
\newcommand{\adv}{\ub \cdot \nab}
\newcommand{\taup}{\tau_\rp}
\newcommand{\tauk}{\tau_k}
\newcommand{\taukN}{\tau_{k,{\rm N}}}

\newcommand{\PikN}{\Pi_{\rf,{\rm N}}}
\newcommand{\nup}{\frac{\mu_\rp}{\taup}}
\newcommand{\Rel}{Re_\lambda}

\def \Aij {A_{ij}}

\newcommand{\lrp}[1] {\left( #1 \right)}
\newcommand{\lrs}[1] {\left[ #1 \right]}
\newcommand{\lra}[1] {\left \langle #1 \right \rangle}

\newcommand {\RKS}[1] {\textcolor{cyan}{#1}}

\def \nnn {\nonumber \\ }
\def \bB {\bf B}
\def \poly {\textit{polymeric }}
\def \newt {\textit{Newtonian }}

\title{The interplay of inertia and elasticity in polymeric flows}

\author{Rahul Kumar Singh and Marco Edoardo Rosti\corresp{\email{marco.rosti@oist.jp}}}
\affiliation{Complex Fluids and Flows Unit, Okinawa Institute of Science and Technology Graduate University (OIST), 1919-1 Tancha, Onna-son, Okinawa 904-0495, Japan}

\begin{document}
\maketitle

\begin{abstract}
Addition of polymers modifies a turbulent flow in a manner that depends non-trivially on the interplay of fluid inertia, quantified by the Reynolds number $Re$, and the elasticity of the dissolved polymers, given by the Deborah number $De$. We use direct numerical simulations to study polymeric flows at different $Re$ and $De$ numbers, and uncover various features of their dynamics. Polymeric flows exhibit {a non-unique scaling of the energy spectrum} that is a function of $Re$ and $De$, owing to different dominant contributions to the total energy flux across scales, {with the weakening of fluid nonlinearity with decreasing $Re$ also leading to the reduction of the polymeric scaling range}. This behaviour is also manifested in the real space scaling of structure functions. We also shed light on how the addition of polymers results in slowing down the fluid non-linear cascade resulting in a depleted flux, as velocity fluctuations with less energy persist for longer times in polymeric flows, {especially at intermediate $Re$ numbers}. These velocity fluctuations exhibit intermittent, large deviations similar to that in a Newtonian flow at large $Re$, but differ more and more as $Re$ becomes smaller. This observation is further supported by the statistics of fluid energy dissipation in polymeric flows, whose distributions collapse on to the Newtonian at large $Re$, but increasingly differ from it as $Re$ decreases. We also show that polymer dissipation is significantly less intermittent compared to fluid dissipation, and even less so when elasticity becomes large. Polymers, on an average, dissipate more energy when they are stretched more, which happens in extensional regions of the flow. However, owing to vortex stretching, regions with large rotation rates also correlate with large polymer extensions, albeit to a relatively less degree than extensional regions.
\end{abstract}

\section{Introduction}

Polymeric flows are very well known to give rise to a wide range of intriguing phenomena. The pioneering work by~\citet{Toms} showed that a small concentration of polymers added to a carrier flow results in the reduction of turbulent drag, which were reviewed in~\citet{Lumley73, Virk75, Toms77} and~\citet{Berman1978}. Ever since this discovery, there have been numerous theoretical~\citep{Tabor86, JKB91, Sreeni00, Lvov04, Procaccia08}, numerical~\citep{Rama03, Rama05, Prasad06, White08, Gillissen08, Hof18, Li2019, Marco23} and experimental~\citep{Toonder97, Smith99, Ptasinski01, Ptasinski03, Dou10} works in an attempt to better understand and highlight its origins and the underlying mechanisms. \citet{Lumley73} and~\citet{Hinch77} showed that dissolved polymers tend to increase the effective viscosity of a solution, however there can be a net reduction in the dissipation of energy in such turbulent, polymeric flows, as was pointed out by~\citet{Sreeni99, Rama05, Prasad06, Prasad10} and~\citet{Zhang10}. This rich, and complex behaviour of polymeric flows is dictated by an interplay of two intrinsic features: inertia and elasticity. Inertia is a property of the Newtonian, carrier flow, whose strength is determined by the non-dimensional Reynolds number $Re$. A large $Re$ is typically indicative of a highly turbulent flow where flow structures span a wide range of scales. Elasticity, on the other hand, is a characteristic of the polymers, that determines their tendency to stretch and is quantified by the non-dimensional Deborah number $De$. A large $De$ means the polymers take longer to relax back to their equilibrium lengths and remain stretched for longer times. 

Experiments by~\citet{Larson92, Shafqeh96, Groisman98} and~\citet{Arratia13}  have shown that flows at small $Re$ develop instabilities when polymers are added in small concentrations. At extremely small $Re$, when fluid non-linearity remains dormant, instabilities driven by purely elastic effects result in a flow state referred to as elastic turbulence (ET) which shows qualitatively similar behaviour to classical high $Re$ turbulence (HIT) in many respects. These include the existence of a chaotic flow state, as illustrated in the works of~\citet{Groisman00, Groisman04} and~\cite{Arratia17}, as well as enhancement of mixing~\citet{Groisman01} and heat transfer~\citet{Li17}. A self-similar distribution of kinetic energy over a wide range of scales was also observed in numerical simulations of ET~\citep{Boffetta08, Berti10, SSR16, Dario19, SinghET23, soligo_rosti_2023a, Giulio24,Kerswell24}. At larger, yet moderate $Re$, it is now known that polymer elasticity and the fluid non-linearity (albeit with a weaker contribution as shown by~\cite{Gillissen21}) conspire to trigger a turbulence-like behaviour much earlier than that in a Newtonian flow~\citep{Dubief13, Hof13, Silva16, Hof23}. Such a flow state at moderate $Re$ is referred to as elasto-inertial turbulence (EIT). Fluid kinetic energy in EIT is also known to exhibit a distinct self-similar, scaling behaviour~\citep{Zhang21, Dubief13, Silva16}. At very large $Re$ numbers, turbulent polymeric flows (PHIT) exhibit a novel self-similar spectrum whose scaling exponents differ from those predicted by the Kolmogorov theory (K41) for classical Newtonian turbulence (HIT)~\citep{Kolm1941, Frisch96}. This was uncovered recently in the experiments of~\citet{Bodenschatz21} as well as in the numerical simulations of~\citet{Marco23}.
 
In this work, we study turbulent polymeric flows in a triperiodic box across a wide range of $Re$ and $De$ numbers, {and show how various aspects of their dynamics transition when moving from low~\citep{SinghET23} to large~\citep{Marco23} Reynolds number.} We show that the addition of polymers results in a {non-unique scaling} of the energy spectrum, that emerges as a result of different underlying dominant flux contributions. The fluxes furthermore reveal that polymers slow down the rate of energy transfer via the fluid non-linear cascade, signalling a weakened fluid non-linear cascade. We then discuss how polymers are, in turn, themselves affected by Newtonian carrier flow. In particular, we show that the polymers stretch in both the extensional and rotational regions of the flow, with the correlation being less for the latter. Both fluid and polymer extensions are directly related to the corresponding dissipations, which show contrasting characteristics. Fluid dissipation shows more large deviations about the mean at small $Re$ and large $De$, while polymer dissipation is less intermittent, and even lesser at large $De$. We structure all of the above as follows: the details of numerical simulations are provided in the immediately following~\cref{sec:numerical} before moving on to the results in~\cref{sec:results}. Finally, we summarise our conclusions in~\cref{sec:conclusions}.

\section{Mathematical model and numerical method}
\label{sec:numerical}

\begin{figure*}
	\centering
	\includegraphics[width=\textwidth]{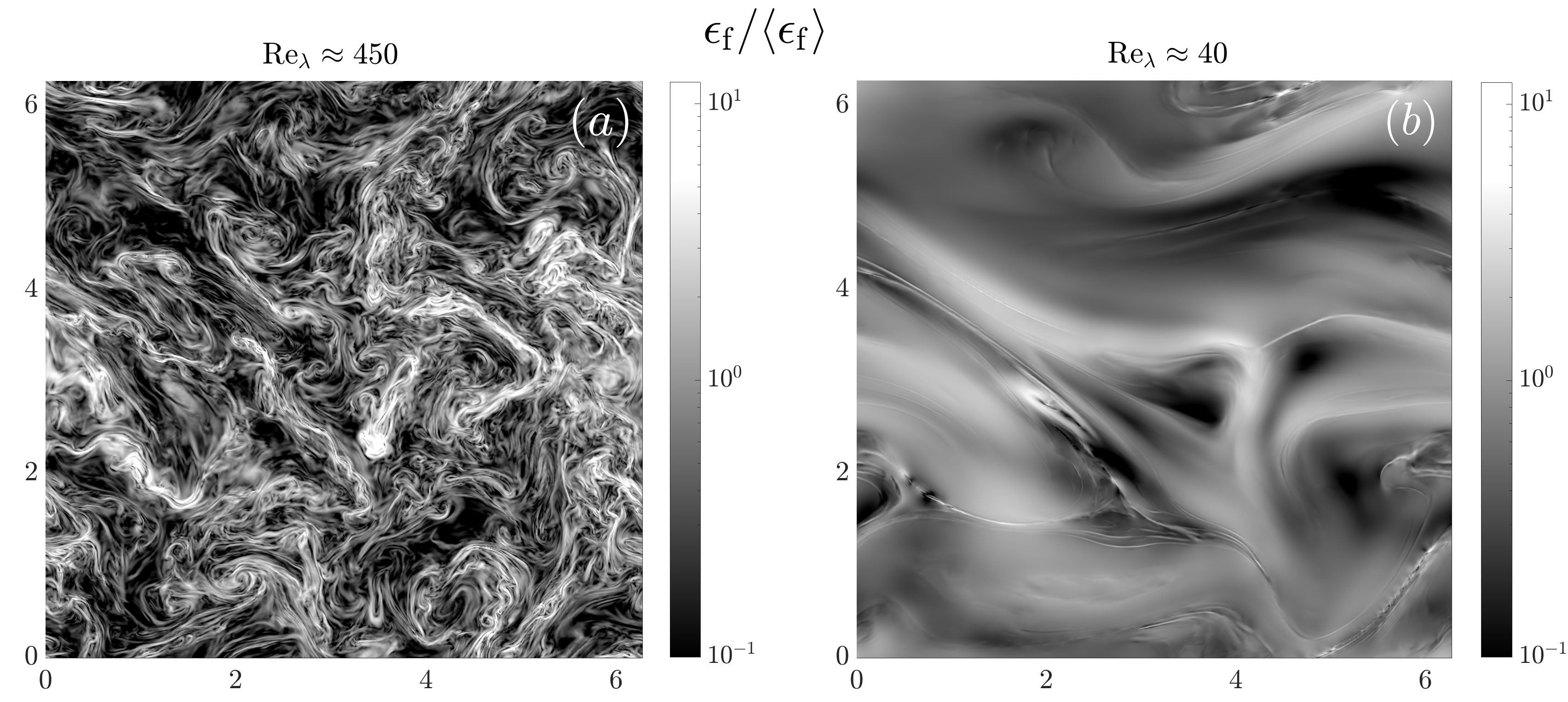}
	\caption{Representative snapshots of normalized fluid energy dissipation rates $\epsilon_{\rf}$ at (a) large and (b) small $ Re $ for $De \approx 1$. Large $\Rel$ polymeric flows exhibit a wide range of flow structures, while the small $\Rel$ flows have only large scale structures.}
	\label{fig:Vis}
\end{figure*}

We arrive at our results via the direct numerical simulations of polymeric flows that comprise an incompressible Newtonian carrier fluid whose dynamics is described by a velocity field ${\bf u}$ ($\nabla \cdot \bu = 0$), with dissolved polymers that are described by a conformation tensor field $\C$.{The trace of this conformation field $\trC = C_{ii} (\bx,t)$  (with $i = 1,2,3$) is an average measure of the squared end-to-end lengths of the polymers}. In particular, we adopt the Oldroyd-B model of polymers, which has often been used to study polymeric flows in various settings in a number of works~\citep{Dario07, Boffetta08, Boffetta12, Prasad12, Dario19, Zhang21, Marco23, aswathy_rosti_2024a}. The flow is described by the simultaneous evolution of the coupled equations:
\begin{subequations}
	\begin{align}
		\rho \left( \pt \ub + \adv \ub \right) &= - \nabla p + \mu_\ri{s} \nabs \ub + \nup  \nabla \cdot \C + \mathbf{F}, \label{NN}  \\
		\pt \C + \adv \C &= \C \nabla \ub + (\nabla \ub)^T \C - \frac{1}{\tau_\rp} (\C -\I),  \label{Conf} 
	\end{align}
\label{eq:MNSE}
\end{subequations}
where $\rho$ is the fluid density, $p$ the pressure, $\mu_\ri{s}$ the dynamic viscosity of the Newtonian fluid (kinematic viscosity $\nu=\mu_\ri{s}/\rho$), $\mu_\ri{p}$ the dynamic polymer viscosity, and $\tau_\rp$ is the unique relaxation time of the polymers. A statistically stationary state is maintained by energy injection via the forcing ${\bf F}$ for which we employ the ABC scheme such that ${\bf F} = \nu [ (A \sin z + C \cos y) \, {\bf \hat{x}} +  (B \sin x + A \cos z) \, {\bf \hat{y}} +( C \sin y + B \cos x) \, {\bf \hat{z}}]$, where $A = B = C = 1$. The total energy injected per unit time $\epsilon_\ri{t}$ is dissipated away by both the Newtonian fluid ($\epsilon_\ri{f}$) and the polymers ($\epsilon_\ri{p}$), such that $\epsilon_\ri{t} = \epsilon_\ri{f} + \epsilon_\ri{p}$.

The flow is characterised by two dimensionless numbers: the Taylor-scale Reynolds number, $\Rel \equiv u_{rms}\lambda/\nu$, where $\lambda \equiv u_{rms}\sqrt{15\nu/\epsilon_\ri{f}}$ is the Taylor length scale and $u_{rms}$ the root mean square velocity, and the Deborah number, $De \equiv \tau_\ri{p}/\tau_\ri{L}$, where $\tau_\ri{L} = L/u_{rms}$ is the large-eddy turnover time, {with $L = 2\pi$ being the scale of the forcing}. {We vary $\Rel$ in the range between$40$ and $450$} and $De \in \{1/9,1/3, 1, 3, 9\}$. Lastly, we fix fluid and polymer viscosities such that $\mu_\ri{s}/\left(\mu_\ri{s}+\mu_\ri{p}\right)=0.9$. {All cases considered in this work are listed in table~\ref{tab:simulations}; note that, the last six simulations with $\Rel \approx 40$ are the same presented by \citet{SinghET23}, while the first six are similar to those discussed by \citet{Marco23}, but repeated here with a different forcing scheme for the sake of consistency with the rest of the study. All the 10 intermediate cases are completely novel, and represent the heart of the present investigation.}

We solve the fluid velocity and polymer conformation tensor field given by~\cref{NN,Conf} on a staggered, uniform, Cartesian grid, using the in-house flow solver \href{https://groups.oist.jp/cffu/code}{Fujin}, which employs a second-order central finite difference scheme for spatial discretisation. Time marching is achieved via a second-order Adams-Bashforth scheme coupled with a fractional step method~\citep{Kim}, except for the non-Newtonian stress term which is advanced in time via the Crank-Nicolson scheme. In the momentum equation we use the Adams-Bashforth scheme scheme for time advancement for all terms except for the polymer contribution, because this can be advanced implicitly with the Crank-Nicolson scheme, since it's updated value is found separately from the additional transport equation. In the polymeric equation, we still use the second-order Adams-Bashforth scheme for all terms, except for the advection term for which we use the high-order Weighted Essentially Non-Oscillatory (WENO) scheme~\citep{Sugiyama11} scheme. A log-conformation formulation is also used for \cref{Conf}, to ensure positive-definiteness of the conformation tensor at all times (see~\citep{Marco18}). Indeed, all these schemes, together with the use of the log conformation formulation for the polymeric equation, allows to overcome the notorious high Deborah numerical instability~\citep{min_yoo_choi_2001a}, without any explicit artificial diffusion term, as discussed in previous papers (see e.g.~\citet{min_yoo_choi_2001a,Sugiyama11,Marco18}. Obtaining incompressible fluid velocity at each time requires solving a Poisson equation for pressure, which is done using a solver based on Fast Fourier Transform (FFT). The solver is parallelized using the domain decomposition library \href{http://www.2decomp.org}{2decomp} and the MPI protocol.

We solve~\cref{eq:MNSE} on a $2\pi \times 2\pi \times2\pi $-periodic domain that is discretized into $N^3 = 1024^3$ points. The smallest spatial scales resolved are equal to $0.05 \eta$ and $\eta$ for the smallest and largest $\Rel$ simulations. We choose two different time-steppings of $\delta t = 1.25\times 10^{-5}$ and $\delta t = 2.5\times 10^{-5}$ for small and large $\Rel$ simulations, which respectively equal $1.5 \times 10^{-3} \tau_\eta$ and $2 \times 10^{-3} \tau_\eta$, where $\tau_\eta$ is the Kolmogorov time-scale. Field data is saved every $2\times 10^4$ time steps in a stationary state for data analysis with the overall data spanning $\approx 6 \tau_L$ for all cases. This means simulating around $54 \tau_{\rm p}$ for the smallest $De = 1/9$,  $6 \tau_{\rm p}$ for $De = 1$ and $\tau_{\rm p}$ for the largest $De = 9$. We have checked by splitting the data set into half for the largest $De$ that the statistics remain unchanged. We show representative, two-dimensional slices of the dissipation fields $\epsilon_{\rf}$  for very large and small $Re$ in~\cref{fig:Vis} while some typical averaged statistical estimates for the simulations are provided in table~\ref{tab:simulations}.

\begin{table}
	\caption{{Details of the numerical simulations considered in the present study. $De$ is the Deborah number, Re$_\lambda$ is the Reynolds number based on $u_{rms} $ and on the Taylor length scale $\lambda$,  $\epsilon_{\rf}, \epsilon_{\rp}$ are the mean fluid and polymer dissipation of energy, $\Omega$ is the mean enstrophy, $\eta$ is the Kolmogorov length scale and $\tau_L$ is the large eddy turnover time scale of the flow.}}
\label{tab:simulations}
\centering
\begin{tabular}{lcccccccccccccccccc}
								 &&               	 &&          				&&     				       		      &&   					       			 &&   		            			      	 &&              								\\
	Re$_\lambda$	&&	$De$	   &&	$u_{rms}$	&&	$\epsilon_{\rm f}$	&&	$\epsilon_{\rm p}$	 &&		$\Omega$				   &&			$\eta$  				  &&			$\tau_L$   				    \\
 								 &&               	 &&          				&&     				       		  	  &&   					       			 &&   		            			      	 &&              								\\
			450			     &&  	--  	    &&     6.6             &&     	  63.0          	     &&   			--		   			  &&      $2.8 \times 10^4$     &&   $3.7 \times 10^{-3}$  	 &&			$9.5 \times 10^{-1}$      	 \\
			470				 &&   1/9    	    &&     6.5	  		   && 		    54.1          		&&       	8.4	       				 &&    	 $2.4 \times 10^4$     &&	$3.8 \times 10^{-3}$	&&		  $9.7 \times 10^{-1}$     	 \\
			583				 &&   1/3    	    &&     7.0	  		   && 	        45.8         	    &&         25.0 	      		   &&      $2.0 \times 10^4$	 &&   $4.0 \times 10^{-3}$	  &&			$9.0 \times 10^{-1}$  \\
			816				 &&     1            &&     6.4     		&& 	         17.1      	    	  &&      	  50.2		 			 &&    	  $7.6 \times 10^3$      &&  $5.1 \times 10^{-3}$	  &&			$9.8 \times 10^{-1}$   	 \\
			713				 &&     3           &&     5.8	 		   &&            14.5         		&&     		44.2  	   			   &&    	$6.4 \times 10^3$      &&	$5.3 \times 10^{-3}$	&&		 $1.1 \times 10^{0}$	     	 \\
			772				 &&     9           &&     6.3	   		   &&            17.4     	  		&&      	52.7 				    &&   	 $7.7 \times 10^3$      &&  $5.1 \times 10^{-3}$	 &&			$1.0 \times 10^{0}$	   	 \\
								 &&               	 &&          				&&     				       		  &&   					       			 &&   		            			      	     &&              \\
			317		    	 &&  	   --       &&     8.0        		 &&      	60.9     		   &&   		--	        			&&   	 $1.4 \times 10^4$       &&   $6.2 \times 10^{-3}$    &&			$9.6 \times 10^{-1}$   \\	
			510				 &&     1     		&&     7.4 				&& 	       16.7     			&&      48.2	     			  &&   		$3.7 \times 10^3$		 &&	   $8.6 \times 10^{-3}$	 &&		 	$1.0 \times 10^{0}$      \\
								 &&              	&&          				&&              	    		 &&            							 &&   		            						  &&                \\	  				
			240			     &&  	 --   	  &&     7.3               &&        83.8               &&   			--             		     &&    	$9.3 \times 10^3$ 		&&	     $9.7 \times 10^{-3}$   &&	$8.6 \times 10^{-1}$   	    \\	 	
			250				 &&   1/9   	 &&     6.9		  		  && 	   61.2     			&&      12.5 	        		     &&  	   $6.8 \times 10^3$ 		&&		$1.0 \times 10^{-2}$   &&  $9.1 \times 10^{-1}$          \\
			293				 &&   1/3   	 &&      6.1 	  		   && 	   26.5     			&&      30.7 	        		    &&   	   $2.9 \times 10^3$		&&		$1.3 \times 10^{-2}$   &&  $1.0 \times 10^{0}$ 	    	   \\
			316				 &&   1      	  &&       5.5	   			&& 	    15.4        		 &&       42.6	        		     &&    		$1.7 \times 10^3$	      &&	$1.5 \times 10^{-2}$	 &&	 $1.1 \times 10^{0}$ 	 	  \\
			312				 &&   3      	 &&      6.3				&&        26.7 				&&      52.5 	        		    &&   	   $3.0 \times 10^3$		&&		$1.3 \times 10^{-2}$   &&  $1.0 \times 10^{0}$ 		\\
			240			    &&   9      	 &&      5.5				&&        27.0      		&&      44.3 		    		    &&   	   $3.0 \times 10^3$		&&		$1.3 \times 10^{-2}$  &&  $1.1 \times 10^{0}$ 	 	   \\
								 &&              &&          					&&              				&&            							&&   		            							&&      	                     \\
			103		   		 &&  	--      &&        6.2      		    &&     	  58.4              &&   	--			    	          &&     	 $1.6 \times 10^3$         	&&     $3.0 \times 10^{-2}$  &&	 $1.0 \times 10^{0}$ 	\\
			166     		 &&   1      	&&       6.1				&& 		  20.3     		    &&      49.3 	       				&&   	   $5.6 \times 10^2$ 		  &&	$3.9 \times 10^{-2}$	&&	$1.0 \times 10^{0}$	 	\\
								&&               &&          					&&   				           &&   				         		   &&   							                &&      						                &&              \\
 			42			     &&  	 --      &&        6.0			     &&        60.9   	    	&&   		--  			          &&        $2.3 \times 10^2$       &&  	 $9.9 \times 10^{-2}$	&&  	$1.0 \times 10^{0}$		\\
			56				&&   1/9       &&    	6.5				  && 	     49.2	          &&   		50.5	                 &&    	    $2.7 \times 10^2$		&&		  $1.0 \times 10^{-1}$	  &&  	  $1.0 \times 10^{0}$	\\
			39				&&   1/3       &&    	5.2			  	  && 	     39.1              &&   	62.3		    		  &&  		 $2.2 \times 10^2$		 &&		   $1.1 \times 10^{-1}$	   &&      $1.2 \times 10^{0}$	\\
			49				&&   1          &&        5.5			   && 	 	   31.2             &&     	 44.7	        		   &&   	   $1.7 \times 10^2$   	   &&		 $1.2 \times 10^{-1}$	&&      $1.1 \times 10^{0}$		\\
			34				&&   3         &&		4.4			      &&          25.5   		   &&      33.9 	  				  &&         $1.4 \times 10^2$ 		 &&	       $1.2 \times 10^{-1}$		&&	  $1.4 \times 10^{0}$\\
			32				&&   9         &&      	4.4				  &&          30.2		       &&      26.8        			     &&      	 $1.7 \times 10^2$	     &&			$1.2 \times 10^{-1}$	&&    $1.4 \times 10^{0}$ \\
								&&              &&          				  &&             				   &&            						  &&   		            &&      	                &&              \\
\end{tabular}
\end{table}

\section{Results}
\label{sec:results}

\subsection{The energy spectra}
\label{sec:Spectra}

\begin{figure*}
	\centering
	\includegraphics[width=\textwidth]{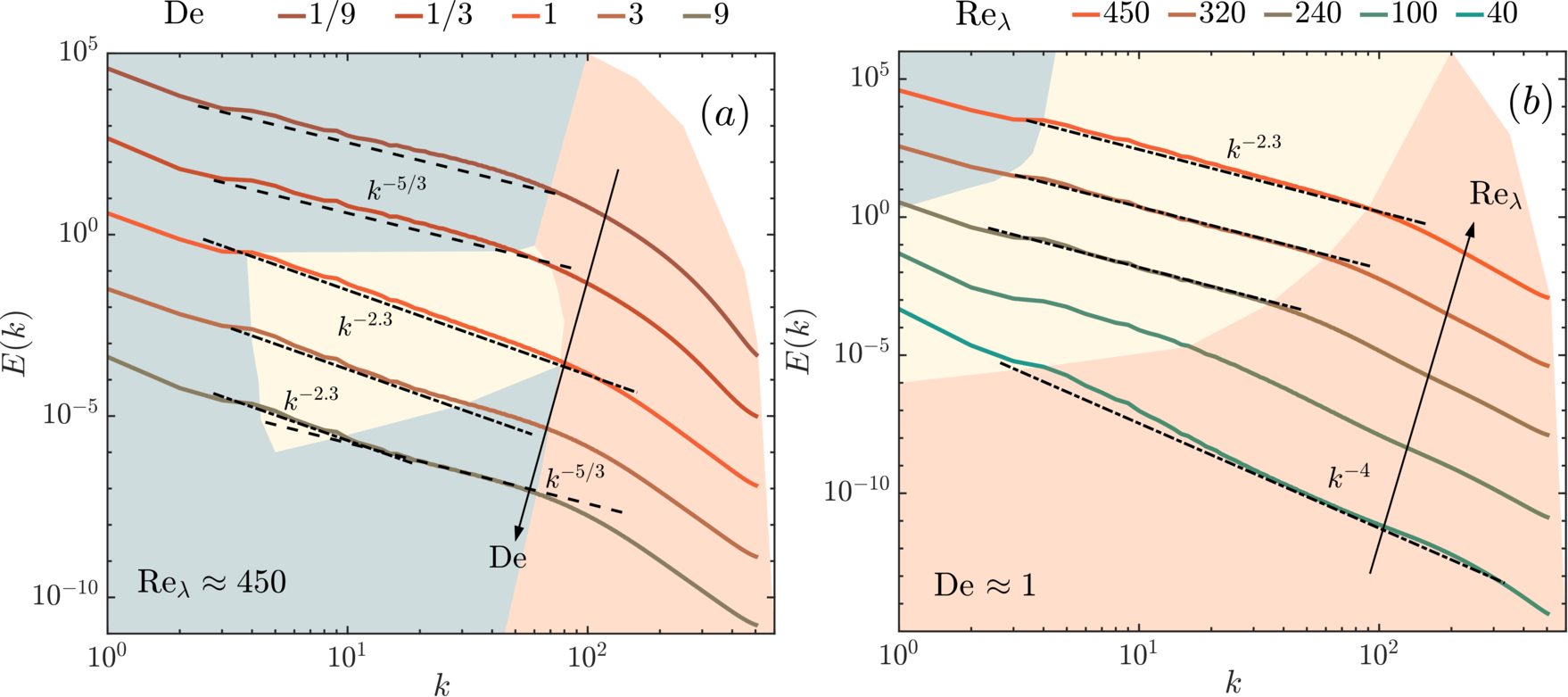}
	\caption{The {non-unique scaling} of the fluid energy spectrum $E(k)$ in polymeric flows at different (a) $De$ and (b) $Re$ numbers. {The spectra have been shifted vertically for visual clarity by factors of powers of 10}. Three distinct scaling regimes (in different shades) have been shown in dashed ($k^{-5/3}$, Newtonian), dash-dotted ($k^{-2.3}$, polymeric) and dotted ($k^{-\gamma}$; $\gamma \geq 4$, smooth) lines. All three regimes coexist when De $\approx 1$. (a) The smallest $De$ has a close to Newtonian behaviour as the elastic effects are minimal. (b) The triple scaling at largest $Re$ gives way to a unique, smooth $k^{-4}$ regime at the smallest $\Rel$.}
	\label{Ek}
\end{figure*}

We begin our study by investigating how energy is distributed across scales in a statistically stationary state in flows of polymeric fluids. The distribution of energy is given by its spectrum $E(k)$ which is a measure of the energy content at scale (mode) $k$ such that the total energy $\mathcal{E} = \langle |\bu|^2 \rangle/2 = \int E(k) dk $. Energy spectrum is one of the most commonly studied quantities in turbulence and was shown by~\citet{Kolm1941} to have a self-similar distribution across scales in (the inertial range of) HIT as $E(k) \sim k^{-5/3}$ (see also~\citet{Frisch96}). We now show in~\cref{Ek} how this self-similar distribution of fluid kinetic energy is modified in PHIT, and its dependence on polymer elasticity $De$ in panel (a) and the intensity of turbulence $\Rel$ in panel (b).

\cref{Ek}(a) shows how the spectrum $E(k)$ changes with $De$ at large $\Rel \approx 450$. When $De$ is small, i.e. in the limit $\tau_{\rp} \to 0$, polymers are stretch minimally, and remain close to their equilibrium unstretched configuration $\C \approx \mathbf{I}$. This is easily seen by multiplying~\cref{Conf} by $\taup$, and taking the limit $\taup \to 0$:
\begin{align}
\taup \lrp{\pt \C + \adv \C} = \taup \lrp{\C \nabla \ub + (\nabla \ub)^T \C} -  (\C -\I)		\overset{\taup \to 0}{\implies} \C -\I = 0.
\end{align}
As the fluctuations in $\C$ are small, the polymer stresses in~\cref{NN} have a minimal contribution: $\nabla \cdot \C \approx 0$. Thus, elastic effects do not play a significant role at small $De$, and the behaviour remains largely Newtonian. This results in the classical K41, Newtonian like scaling $E(k) \sim k^{-5/3}$ in~\cref{Ek}(a). (Henceforth, we refer to this classical scaling range as the \newt regime.) This intermediate, \newt scaling range is of course followed by the dissipation range where the spectrum shows an exponential decay, a consequence of velocity fields being analytic.

As $De$ is increased, polymers begin to stretch to longer lengths so that $\C$ now ventures far from $\I$ and has rather large spatio-temporal fluctuations. The resulting stretching and relaxation of polymers generates large non-Newtonian stresses (given by $\lrp{\mu_\rp/\taup} \nabla \cdot \C$) on the carrier fluid, especially for a large enough elasticity as $De \approx 1$. The resulting dominant elastic effects alongwith a suppressed fluid non-linearity (see~\cref{sec:Fluxes} for detailed discussion) means that the energy spectrum is modified to $E(k) \sim k^{-2.3}$ at large $\Rel$, as was also observed by~\citet{Zhang21} and~\citet{Marco23}. (We refer to this as the \poly range henceforth.) Viscous effects begin to dominate beyond this \poly range and the fluid non-linearity ceases to have any contribution. The polymeric stresses, however, are still active in this range due to the small scale fluctuations in polymer lengths which, in turn, create fluctuations in the 	fluid velocity. Consequently, the spectrum exhibits a steep power-law decay $E(k) \sim k^{-\delta}$ with $\delta \geq 4$, rather than an exponential fall-off in the \text{smooth} dissipative range. Thus, the elasticity of polymers results in a dual scaling behaviour of the energy spectrum $E(k) \sim k^{-\delta}$ in PHIT, especially for $De \approx 1$ (see~\cref{Ek}(a)), with the exponent $\delta \approx 2.3$ in the intermediate, polymeric range of scales ($6 \lesssim k \lesssim 60$) and $\delta \gtrsim 4$ for $k \gtrsim 100$ in the smooth but steep dissipative .
   
On a further increase in polymer elasticity, the back reaction of polymers on the Newtonian carrier flow begins to diminish. This is because polymers with very large relaxation times respond to the background flow after a large time lag, thereby staying in a stretched configuration for long times. {This means fluctuations in polymer lengths are suppressed at large $De$ as polymer lengths do not change appreciably in space. As a result, their gradients also remain typically small thereby weakening the elastic stresses $\nabla \cdot \C$ on an average.} This is coupled with the fact that a large factor of $\tau_{\rp}$ further weakens the feedback of polymer stresses on the carrier flow. Thus, \newt scaling is recovered for large $De$, and is seen in the re-emergence of $E(k) \sim k^{-5/3}$ regime at $De \approx 9$. Indeed, at even larger $De$ we have an intermediate range of scales with completely \newt scaling behaviour~\citet{Marco23}. Again in the deep dissipation range, small fluctuations in the polymer lengths result in fluctuations of the flow field, which are rough at the sub-leading order with a steep power-law decay of the spectrum as $E(k) \sim k^{-4}$. 

\cref{Ek}(b) shows the dependence of the energy spectrum on $\Rel$, where we plot $E(k)$ for a fixed polymer elasticity of $De \approx 1$. The top curve is just the spectrum at $\Rel \approx 450$ and $De \approx 1$. Now, it is well known that with decreasing $\Rel$ in HIT the fluid non-linearity becomes increasingly less important (with respect to viscous dissipation) causing the inertial range to shrink (and the viscous dissipation range to widen). In PHIT, a decreasing $\Rel$ entails that now the intermediate \poly range shrinks as seen from~\cref{Ek}(b). This is a consequence of the fluid inertia weakening even further in the presence of polymers (see~\cref{sec:Fluxes} for detailed discussion). A further weakened fluid inertia means that viscosity becomes important at even larger scales. The dissipative range thus widens, and the steep power-law fall-off begins at even smaller $k$ as $\Rel$ decreases. In fact, at $\Rel \approx 40$, the \poly regime is completely absent and the smooth dissipation range spans the entire range of scales (see also~\cite{SinghET23}). This is indeed very different from Newtonian turbulence, where a decreasing $\Rel$ results in a shrinking inertial range that completely disappears as $\Rel \to 0$ and the spectrum decays exponentially. Polymeric turbulence, on the other hand, shows a transition from $E(k) \sim k^{-2.3}$, as $\Rel$ decreases, to a steeper power-law spectrum of $E(k) \sim k^{-4}$ as $\Rel \to 0$ for large $De$.

Next, we discuss how different dominant contributions lead to this non-unique scaling behaviour in polymeric flows, as a function of $\Rel$ and $De$.

\subsection{The flux contributions}
\label{sec:Fluxes}

The co-existence of various scaling regimes of the energy spectrum, as discussed in~\cref{sec:Spectra}, can be better understood by looking at the dominant energy transfer mechanisms across scales {(see also~\citet{Casciola2007} for a similar discussion on spectral decomposition with a slightly different approach).} We obtain these dominant contributions from~\cref{NN} by computing its Fourier transform and its complex conjugate,
\begin{subequations}
	\begin{align}
		\partial_t \uh_i(\bk) + \mathcal{N}_i(\bk) &=   i k_i \hat{p} - \nu k^2 \uh_i (\bk) + \frac{\mu_\rp}{\rho \tau_\rp} ik_j  \ch_{ij} + \hat{F}_i(\bk) \	, \quad  k_i \uh_i = 0; 	\label{eq:FT}					\\
		\partial_t \uh^\dag_i(\bk) + \mathcal{N}^\dag_i(\bk) &=   -i k_i \hat{p}^\dag - \nu k^2 \uh^\dag_i (\bk) - \frac{\mu_\rp}{\rho \tau_\rp} ik_j  \ch^\dag_{ij} + \hat{F}^\dag_i(\bk), 	\label{eq:FTC} 
	\end{align}
\end{subequations}
where the hat $\hat{\cdot}$ denotes the Fourier transformed variables, $\cdot^\dag$ denotes the complex conjugate, and $\mathcal{N}_i(\bk)$ is the Fourier transform of the fluid non-linear term. Summation over repeated indices is implied. The energy equation for any $k$-th mode can be obtained by multiplying~\cref{eq:FT} by $u^\dag_i(\bk)$, and~\cref{eq:FTC} by $u_i(\bk)$, and adding the two together:
\begin{align}
	\pt \mathcal{E}(\bk) +  \Re \lrp{\uh^\dag_i(\bk) \mathcal{N}_i (\bk)} =  - 2 \nu k^2 \mathcal{E}(\bk)  + \frac{\mu_\rp}{\rho \tau_\rp} \Im \lrp{\uh^\dag_i(\bk) k_j  \ch_{ij} } + \Re \lrp{\uh^\dag_i(\bk) \hat{F}_i(\bk)},
\end{align}
where $\mathcal{E}(\bk) =  \lvert \uh_i (\bk) \rvert^2 /2 $ is the energy in mode $\bk$, the pressure term vanishes away due to incompressibility, and $\Re \lrp{\cdot}$ and $\Im \lrp{\cdot}$ denote the real and imaginary parts respectively. The total energy flux from modes $k < K$ to those $k > K$ can be obtained by integrating from $0$ to $K$:
\begin{align}
&	\pt \int_{0}^{K} d\bk \ \mathcal{E}(\bk) +  \int_{0}^{K} d \bk \ \Re \lrp{\uh^\dag_i(\bk) \mathcal{N}_i (\bk)} = \nonumber \\
& -  \nu \int_{0}^{K} d\bk \ k^2 \lvert \uh_i(\bk) \rvert ^2 + \frac{\mu_\rp}{\rho \tau_\rp} \int_{0}^{K} d \bk \ \Im \lrp{ \uh^\dag_i(\bk) k_j  \ch_{ij} }  + \int_{0}^{K} d \bk \  \Re \lrp{\uh^\dag_i(\bk) \hat{F}_i(\bk)}. 					\label{eq:Budget}
\end{align}
\begin{figure}
	\centering
	\includegraphics[width=\textwidth]{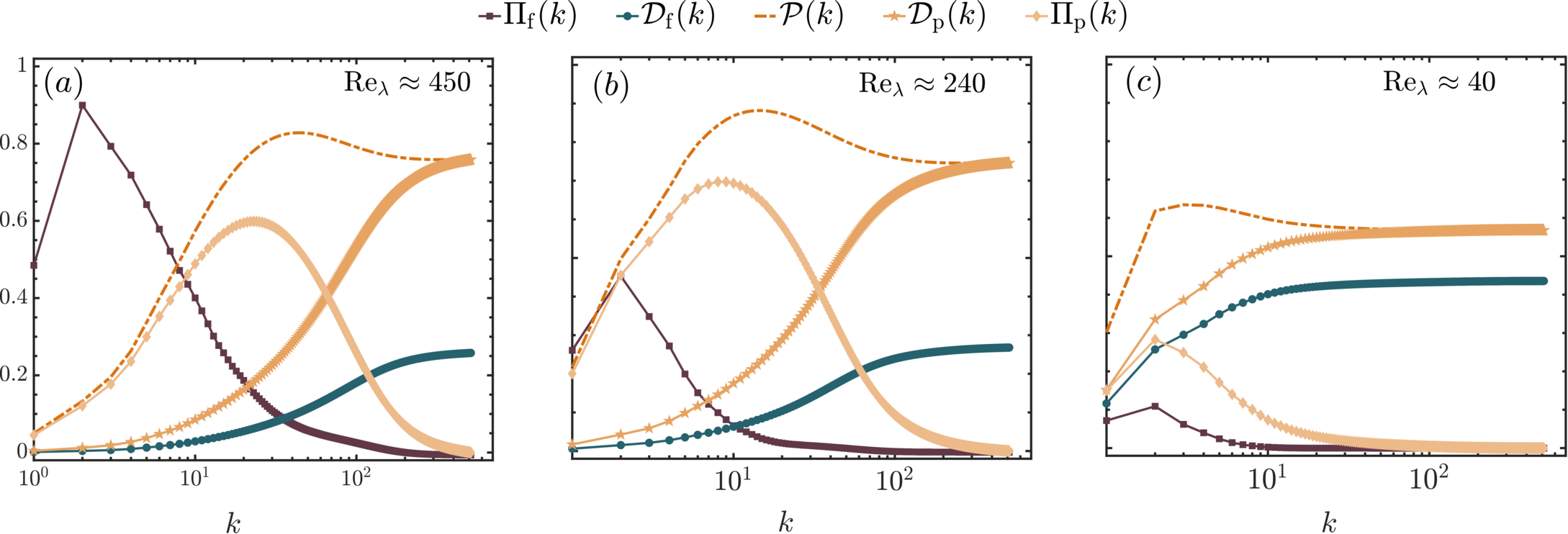}
	\caption{Normalized flux contributions at (a) Re$_\lambda \approx 450$, (b) $240$, and (c) $40$ for flows with De $\approx 1$. {The polymeric contribution $\calp$ is split into flux $\Pi_\ri{p} $ and dissipation $\caldp$ contributions as $ \calp = \Pi_\ri{p} + \caldp$ as in~\citet{Marco23}.} (a) At large $\Rel$, three distinct regimes are determined by different dominant contributions: large scales are dominated by the fluid non-linear flux $\Pi_\ri{f}$, intermediate scales by the polymer flux $\Pi_\ri{p}$, and small scales by the polymer dissipation $\mathcal{D}_\ri{p}$. (b) At moderate $\Rel$, the fluid-nonlinearity $\Pi_{\rf}$ is weakened and comprises two distinct regimes dominated by $\Pi_\rp$ and $\caldp$. (c) At extremely small $\Rel$, only the dissipative terms $\caldf$, $\caldp$ remain important. {All terms are normlised by $\epsilon_{\rm t}$.}}
	\label{fig:Flux}
\end{figure}
In a statistically stationary state, the total energy in any set of modes is a constant, so that $\partial_t \int_{0}^{K} d\bk \ \mathcal{E}(\bk) = 0$. Choosing suitable symbols for the remaining integrals, and using isotropy to argue that their dependence is only on $K = \lvert {\bm K}  \rvert$, we have:
\begin{align}
	\Pi'_{\rf}(K) = - \mathcal{D}'_{\rf}(K) - \mathcal{P}'(K) + \mathcal{F}(K),
\end{align}
where $\Pi'_{\rf}$, $\mathcal{D}'_{\rf}$, $\mathcal{P}'$, and $\mathcal{F}$ are the contributions from fluid non-linearity, fluid dissipation, polymer stresses, and the external forcing, respectively. Since the forcing $\hat{F}_i$ is applied at only $K = 1$, we have that $\mathcal{F}(K)$ is a constant for any $K > 1$:
\begin{align}
	\Pi'_{\rf}(K) + \mathcal{D}'_{\rf}(K) + \mathcal{P}'(K) = \mathcal{F}(K)  = \epsilon_{\rm t},
\end{align}
where $\epsilon_t$ is the rate of total energy injection into the system by the forcing $\mathbf{F}$. Thus, upon normalization by $\epsilon_{\rm t}$, we obtain:
\begin{align}
	\Pi_{\rf}(K) + \mathcal{D}_{\rf}(K) + \mathcal{P}(K) = 1,
\end{align}
where $\Pi_{\rf} = \Pi'_{\rf}/\epsilon_{\rm t} $, $\mathcal{D}_{\rf} = \mathcal{D}'_{\rf}/\epsilon_{\rm t}$, $\mathcal{P} = \mathcal{P}'/\epsilon_{\rm t}$ (see also \citet{abdelgawad_cannon_rosti_2023a}). We show the curves for all the three contributions, $\Pi_{\rf}$, $\mathcal{D}_{\rf}$, and $\mathcal{P}$, as a function of $\Rel$ in~\cref{fig:Flux}, for a polymer elasticity of $De \approx 1$. Following~\citet{Marco23}, we partition the polymeric contribution into flux ($\Pi_{\rp}$) and dissipative ($\mathcal{D}_{\rp}$) terms as:
\begin{align}
	\mathcal{P}(K) = \Pi_{\rp}(K) + \mathcal{D}_{\rp}(K) \ ,  \quad \textrm{where} \quad  \mathcal{D}_{\rp}(K) \equiv (\epsilon_{\rp}/\epsilon_{\rf}) \mathcal{D}_{\rf}(K).
\end{align} 
Such a definition is motivated by the requirement that at small scales (or large $k$) the polymeric term must get all its contribution from dissipative effects. 
Indeed, the full polymeric contribution $\mathcal{P}$ captures both the dissipative and flux contributions. We note that a purely dissipative part $\mathcal{D}_{\rm p}$ of $\mathcal{P}$ must be monotonically growing in $k$, thus we choose to model $\mathcal{D}_{\rm p}$ in the very simple and naive way by prescribing the same scale dependence as the fluid dissipation $\mathcal{D}_{\rm f}$, which we already know grows monotonically in $k$. We tried different prescriptions for $\mathcal{D}_p$ under the constraint of $\mathcal{D}_{\rm p} = 0$ at $k=0$, $\mathcal{D}_{\rm p} = \epsilon_{\rm p}$ at $k=k_{max}$ with a monotonic growth. The results do not change qualitatively but only quantitatively, in the sense that the boundaries of the different scaling ranges shift only marginally. Since these changes remain limited, the estimates for where different flux contributions crossover is reasonably robust. We plot the contributions $\Pi_{\rp}$ and $\mathcal{D}_{\rp}$ thus obtained in~\cref{fig:Flux}.

At large $\Rel$, \cref{fig:Flux}(a) shows that the fluid non-linearity $\Pi_{\rf}$ has the dominant contribution to the total energy balance at large scales (i.e. for $k \leq 4$). This gives the \newt $k^{-5/3}$ scaling at the large but slim band of scales in~\cref{Ek}. Away from this range, i.e. for $k \gtrsim 10$, the polymeric flux contribution $\Pi_{\rp}$ begins to dominate. This dominant non-Newtonian flux results in a \poly scaling regime with $E(k) \sim k^{-2.3}$ upto $k \approx 70$ in~\cref{Ek} (see also~\citet{Marco23}). At yet smaller scales, i.e. for $k \gtrsim 200$, fluid ($\mathcal{D}_{\rf}$) and polymer ($\mathcal{D}_{\rp}$) dissipation remove energy from the flow rapidly. The polymer contribution $\Pi_{\rp}$, has a small, sub-dominant yet non-zero contribution in this range. (This aligns with our assertion of small scale fluctuations in polymer lengths in~\cref{sec:Spectra}.) This sub-dominant contribution results in a fluid velocity spectrum given by a steep power-law decay at large $k$, in a manner similar to that in ET at very small $\Rel$ (see~\citet{SinghET23}). Now, as $\Rel$ is decreased to a moderate value of $\Rel \approx 240$, the fluid non-linearity is weakened evidently further as seen in~\cref{fig:Flux}(b), and the flux of energy is primarily via the fluid-polymer interactions $\Pi_{\rp}$. However, this range is rather limited as a smaller $\Rel$ also means that viscous effects become important at relatively smaller $k$. Thus, we have a restricted \poly and a wider $k^{-4}$ regime at $\Rel \approx 240$ in~\cref{Ek}(b). Finally, for very small $Re_\lambda \approx 40$ in~\cref{fig:Flux}(c), $\mathcal{D}_{\rp}$ and $\mathcal{D}_{\rf}$ are always large while $\Pi_{\rf}$ and $\Pi_{\rp}$ remain sub-dominant. In fact, $\Pi_{\rf}$ at such small $\Rel$ is expected to be dormant and indeed shows an exponential fall off in wave-numbers $k$ in~\cref{fig:Flux2}(c). $\Pi_{\rp}$ instead decays only as a steep power-law, and forces fluctuations in the velocity field that result in the steep, and extended $k^{-4}$ range in~\cref{Ek}(d) . 

\subsection{Slowing down the energy cascade}
\label{sect:Time}

\begin{figure}
	\centering
	\includegraphics[width=\textwidth]{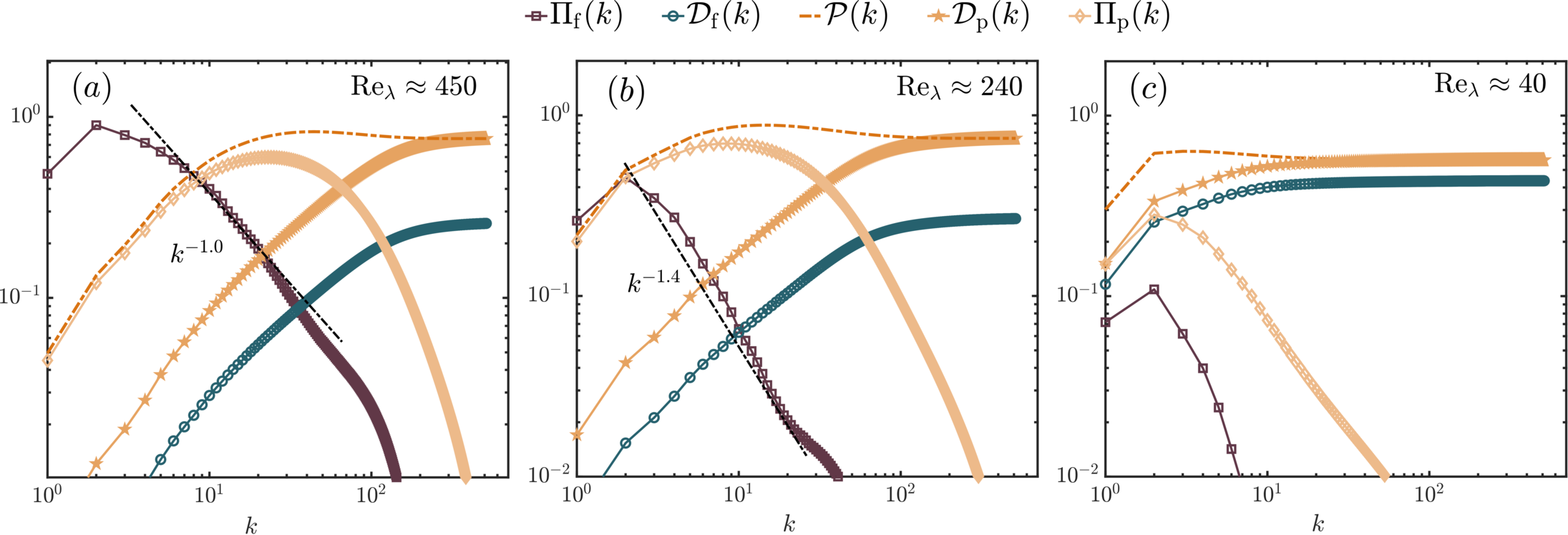}
	\caption{The scaling behaviour of flux contributions obtained by replotting~\cref{fig:Flux} on a log-log scale. The approximate scaling forms are shown as dash-dotted lines. The fluid flux $\Pi_{\rf}$ decays as a power-law in the presence of polymers at large to moderate $\Rel$, implying a slow and weak fluid non-linear cascade.}
	\label{fig:Flux2}
\end{figure}

While the above discussion in~\cref{sec:Spectra} on flux contributions sheds light on the non-unique scaling nature of the energy spectra, there is more that remains in hiding in~\cref{fig:Flux}. We uncover this by plotting the flux contributions on a log-log scale in~\cref{fig:Flux2} which clearly reveals the $k$-dependence of the flux contributions. We are particularly interested in the fluid $\Pi_{\rf}$ and polymeric $\Pi_{\rp}$ fluxes.

\cref{fig:Flux2}(a) shows that at large $\Rel$ our \poly regime $E(k) \sim k^{-2.3}$ is marked by a fluid non-linear flux decaying as $\Pi_{\rf} \sim k^{-1.2}$, while $\Pi_{\rp}$ remains almost constant. Now, $E(k)$ and $\Pi_{\rf}$ can be related to each other using the following simple arguments. $\Pi_{\rf}$ is the average rate of net energy transfer through a scale $k$. This means velocity fluctuations $\uk$ \RKS{upto the scale $k$} transfer their energy to smaller scales in some characteristic time $\tauk \sim (k \uk)^{-1}$. In HIT, this characteristic time-scale is equal to the typical lifetime of an eddy within the K41 phenomenology. This is because energy flux $\PikN$ in the inertial range is a constant, i.e. $\PikN = \epsilon $, with $\epsilon$ being the average rate of energy dissipation. Therefore, we have 
\begin{align}
	\PikN \sim \frac{\uk^2}{\tauk} \sim k \uk^3 \sim \epsilon	\implies  \uk \sim k^{-1/3} 
\end{align}
in the inertial range of scales. This means that the time to transfer energy to smaller scales and the typical lifetime of eddies are respectively estimated as 
\begin{align}
	\taukN \sim \frac{\uk^2}{\PikN} \sim \frac{k E_k}{k^0} \sim k^{-2/3} \ \ ; 	\qquad    \taukN  \sim   (k \uk)^{-1} \sim (k k^{-1/3})^{-1} \sim  k^{-2/3}.	\label{eq:HITime}
\end{align}
%
%
In HIT, these time scales are identical as they both are a consequence of a constant inertial range energy flux  $\PikN$. This, however, is no longer true in PHIT where the fluid non-linear flux $\Pi_{\rf}$ is no longer a constant, but is rather a function of the scale $k$ (and therefore is no longer the invariant that determines the scaling form of the velocity fluctuations). 
{For the remainder of this section, we refer to $De = 1$ as it shows maximal deviation from a Newtonian behaviour. We can still estimate the eddy turnover time-scale in PHIT as 
\begin{align}
	\tauk \sim  (k \uk)^{-1} \sim \lrp{k \sqrt{kE_k}}^{-1} \sim k^{-0.3}. 			\label{eq:tauk1} 
\end{align}
So, the typical lifetimes of fluctuations in polymeric turbulence are larger compared to those in Newtonian turbulent flows as $\tauk$ now falls off slower with $k$. Thus, energy is transferred at a slower rate to smaller scales of the fluid in the presence of polymers. With this knowledge, we can now determine the fate of the fluid nonlinear flux $\Pi_{\rf}$ via the relation $\tauk \sim  \uk^2/\Pi_{\rf}$ as:
\begin{align}
	\Pi_{\rf}  \sim  \frac{\uk^2}{\tauk}  \sim \frac{k E_k}{k^{-0.3}} \sim k^{-1.0} 	\label{eq:Pif} 
\end{align}
Clearly, the scale-by-scale energy transfer via the nonlinear cascade is weakened in polymeric turbulence by the virtue of a smaller proportion of the (total averaged) fluid kinetic energy $\uk^2/u^2 \sim k^{-1.3}$ being passed onto the larger $k$ at a smaller rate $\tauk/\tau_L \sim k^{-0.3}$, compared to a Newtonian fluid. The estimated scaling behaviour of $\Pi_{\rf} \sim k^{-1}$ is indeed found to hold in Fig~\ref{fig:Flux2} for the range $k \in [8,40]$ which is separated from both the forcing effects (acting at $k=1$) and fluid dissipation (which becomes dominant around $k \approx 40 $). (Recall, however, that the energy spectrum scaling $E(k) \sim k^{-2.3}$ holds for the range  $k \in [4,40]$.) This depleted fluid non-linear flux is accompanied by a finite rate of energy transfer to the dissolved polymers. The typical time-scale of this transfer can be estimated by making a crude assumption of a constant polymeric flux, i.e. $\Pi_{\rp} \approx k^0$, in $k \in [10,70]$ from~\cref{fig:Flux2}(a). This assumption then yields the scale-by-scale energy transfer rate to polymers as:
\begin{align}
	\tauk \sim \frac{\uk^2}{\Pi_{\rp}} \sim \frac{k E_k}{k^0} \sim k^{-1.3}.			\label{eq:Pip}
\end{align}
This shows that the remaining energy of fluctuations, that wasn't handed down to the smaller active scales of motion of the carrier fluid by the nonlinear flux, is transferred to the polymers, on an average, at a much faster rate $\tauk \sim k^{-1.3}$.
Now, we expect this relation to hold only approximately for a restricted range of $k \in [10,40]$ as $\Pi_{\rp}$ is not exactly constant in this range.}
 \begin{figure}
 	\centering
 	\includegraphics[width=\textwidth]{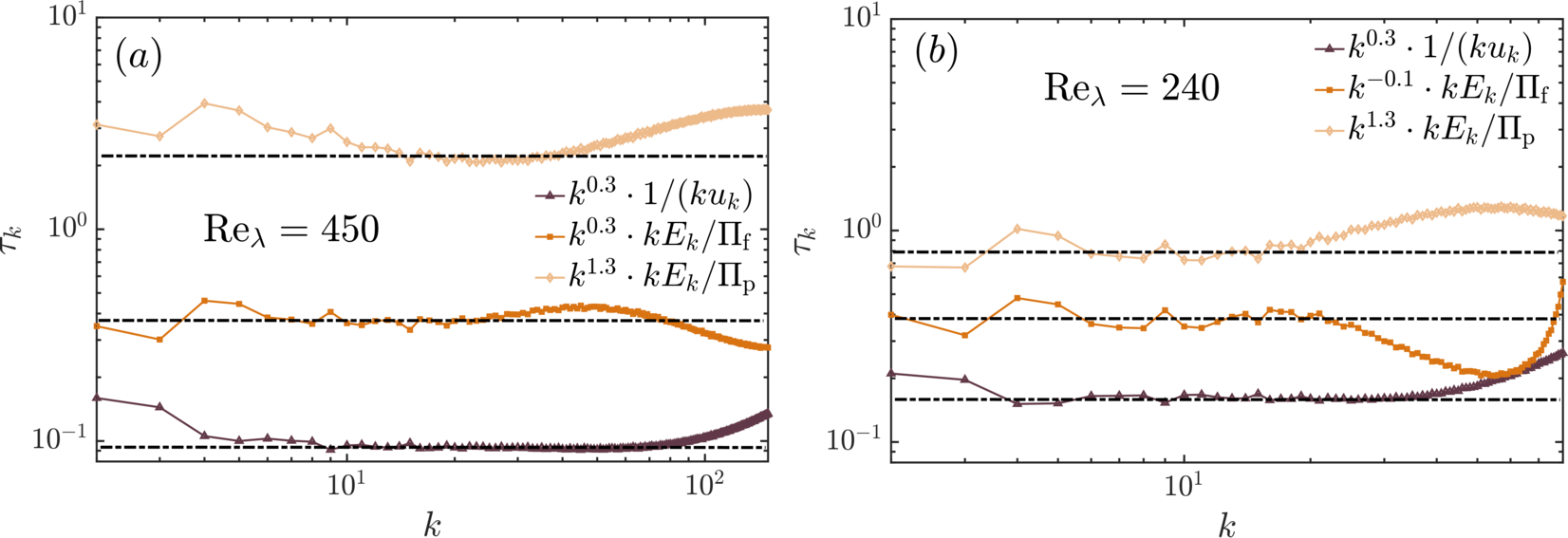}
 	\caption{{The plots of various timescales discussed in section~\ref{sect:Time} for (a) Re$_\lambda$ = 450 and (b) 240, at $De = 1$. We show the compensated plots for the eddy turnover time (triangles) and the time obtained from the nonlinear (squares) and polymer fluxes.}}
 	\label{fig:TT}
 \end{figure}
{ With these caveats in mind, we now plot the compensated time scales in Fig.~\ref{fig:TT}(a) for Re$_\lambda = 450$. The relevant ranges are marked by the flattening of the curves which are computed using the relations~\ref{eq:tauk1},~\ref{eq:Pif} and~\ref{eq:Pip}. Of these, the first two are the estimates of eddy-turnover times/ lifetime of velocity fluctuations in polymeric turbulence. Now, the relations~\ref{eq:tauk1} and~\ref{eq:Pif} must yield the same estimates of $\tauk$ upto some constant of $\mathcal{O}(1)$. (To obtain $\tauk$ from the second relation~\ref{eq:Pif} we use the estimate $\Pi_{\rf} \sim k^{-1}$.) This is indeed found to be consistent in Fig.~\ref{fig:TT}(a) where we show these two estimates as curves with triangle and square markers, respectively. And while these curves do not stay flat over a wide range, it is certainly true for the expected range $k \in [8,40]$. Lastly, the typical time scales of energy transfer to the polymeric mode is given by the compensated curve in diamond markers which corresponds to the relation~\ref{eq:Pip}. In confirmation to our expectations, we indeed find that this curve is (approximately) flat in the range $k \in [10,40]$ in Fig.~\ref{fig:Flux2}(a)). So, it is indeed the case that polymers extract energy at a faster rate from turbulence than the fluid nonlinearity transfers the remaining to smaller scales, on an average resulting in a weakened cascade/fluid nonlinearity. }

{We now turn our attention to when Re$_\lambda = 240$. At the outset, we notice from Fig.~\ref{fig:Flux2}(b) that there is a rather very small set of scales that lies between the forcing scale ($k=1$) and the scale where fluid dissipation becomes important ($k = 10$). Naturally, it cannot be expected that the lifetime of fluctuations $\tauk$ is reliably estimated by relating it to the nonlinear flux via $\Pi_{\rf} \sim \uk^2 /\tauk $, as this relation does not account for losses due to viscous dissipation. However, it still estimates the rate at which energy is transferred to smaller scales. Noting that $\Pi_{\rf}$ still has a power-law behaviour of $\Pi_{\rf} \sim k^{-1.4}$ in $k \in [3,12]$ and $\Pi_{\rp} \approx constant$ in this range as well, we can now estimate the various time-scales at Re$_\lambda = 240$ as: 
\begin{align}
	\tauk  \sim  (k \uk)^{-1}  \sim k^{-0.3} \ ; 	\quad 		\tauk  \sim  \frac{kE_k}{\Pi_{\rf}} \sim k^{0.1} \ ;
	\quad 		\tauk \sim \frac{kE_k}{\Pi_{\rp}} \sim k^{-1.3}.
\end{align}
We plot the corresponding compensated curves of these transfer time-scales in Fig.~\ref{fig:TT}(b).
Consistent with our expectations, all of these curves are fairly constant in the rather small desired range $k \in [4,10]$.  }

In a nutshell, the effect of the addition of polymers is to slow down the cascade process which we quantify using scaling arguments the knowledge of flux contributions, and the fluid energy spectra. In the next section, we discuss how energy is distributed in the polymeric scales of motion via the polymer energy spectrum using data as well as scaling arguments.

\subsection{The polymer spectra}
\label{sec:Polyspec}
\begin{figure*}
	\centering
	\includegraphics[width=\textwidth]{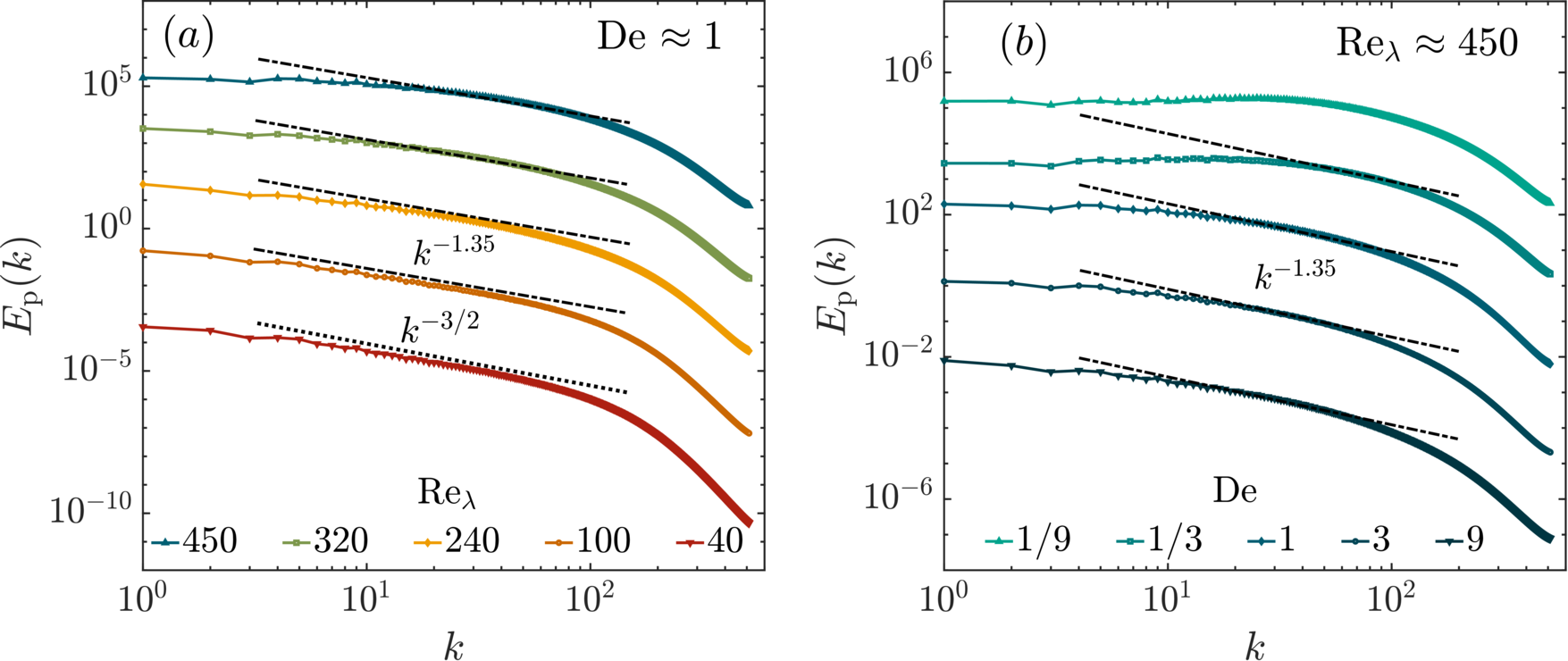}
	\caption{The dependence of the polymer energy spectrum $E_\rp(k)$ on (a) $\Rel$ (shown for $De \approx 1$), and on (b) De (shown for Re$_\lambda \approx$  450). The two different scaling regimes with exponents $-1.35$ (at large Re) and $-1.5$ (at small Re) are shown in dash-dotted and dotted lines, respectively. (b) Small De flows at large Re show a close to Newtonian behaviour and $E_\rp(k)$ remains devoid of any scaling form. As De $\gtrsim 1$, the expected scaling with exponent $-1.35$ begins to appear. {The spectra are shifted vertically for visual clarity by factors of powers of 10.}}
	\label{fig:Epk}
\end{figure*} 
The distribution of energy across polymeric scales, in a statistically stationary state, can be obtained by computing the polymer energy spectrum $E_\rp (k)$ in analogy with the fluid energy spectrum $E (k)$. To do so, {we start by defining the polymer energy spectrum in a sense similar to that of~\citet{Casciola2007,Doering11,Bos16}}. The measure of average total polymer energy $ \mathcal{E}_\rp$ is given by the $\trC$ as:
\begin{align}
 \mathcal{E}_\rp = 	\frac{\mu_\rp}{2 \tau_\rp} \lra{\trC} = \frac{\mu_\rp}{2 \tau_\rp  V} \int_{V} C_{ii}(\bx) d\bx  = \frac{\mu_\rp}{2 \tau_\rp  V} \int_{V} B_{ij}(\bx) B_{ji}(\bx) d\bx  ,
\end{align}
where $\bB$ (with components $B_{ij}$) is the positive-definite square root of the conformation tensor ${\bf C}$, i.e. $C_{ij} (\bx,t) = B_{ik} (\bx,t) B_{kj} (\bx,t)$. One can now define the polymer energy spectrum $E_\rp(k)$ by employing Parseval's relation
\begin{align}
	\mathcal{E}_\rp = \frac{\mu_\rp}{2 \tau_\rp  V} \int_{V} B_{ij}(\bx) B_{ji}(\bx) d\bx  &=  \frac{\mu_\rp}{2 \tau_\rp  V} \int_{\mathbb{R}^3} B_{ij}(\bk) B_{ji}(-\bk) d\bk =   \int_0^{\infty} E_\rp(k) \ dk \nnn 
\text{where} \quad 	E_\rp(k) &= \frac{\mu_\rp}{2 \tau_\rp  V} \int_{|\bk| = k} B_{ij}(\bk) B_{ji}(-\bk) \ d\Omega_k,
\end{align}
where $V$ is volume of integration and the last integral is defined over a spherical shell of radius $k$. 

Now, the scale-by-scale energy $(\mu_\rp/2\taup)C_k$ can then be estimated as the energy contained in the modes $\in \lrs{k, k+dk}$:
\begin{align}
	\frac{\mu_\rp}{2 \tau_\rp}  C_k  =   E_\rp(k) dk,			\label{eq:Epk}
\end{align}
in analogy with the classical
\begin{align}
	\frac{1}{2}  u_k^2  =   E_k (k) dk.			\label{eq:Ek}
\end{align}
Thus, it is easy to see using dimensional arguments that $C_k  \sim k E_\rp(k)$, similar to $u_k^2  \sim k E_k(k)$. We now use this relation to estimate $E_\rp(k)$ by relating it with $E(k)$. 

Let us begin by looking at the $\Pi_\rp$ curves in~\cref{fig:Flux} (or~\cref{fig:Flux2}). For large $\Rel$, the flux contribution $\Pi_{\rp}$ remains approximately constant for $k \in \lrs{20,60}$. Therefore, 
\begin{align}
	\nup  k \uk \ck \sim k^0  \implies  \ck \sim \lrp{k\uk}^{-1}  \sim  k^{-0.35}.				\label{eq:Ck}
\end{align}
This immediately implies, using~\cref{eq:Epk}, that $E_\rp(k) \sim k^{-1} C_k \sim k^{-1.35}$ for $k \in \lrs{20,60}$. Indeed, we find this to be consistent with the data from simulations plotted in~\cref{fig:Epk}(a).  
At $\Rel \approx 240$, this scaling range of $E_\rp(k)$ shows a slight shift to larger scales, in correspondence to a similar shift of constant $\Pi_{\rp}$ in~\cref{fig:Flux2}. 
{We point out here that in contrast to the behaviour of  $E(k)$, the scaling range for $E_\rp (k)$ appear cleaner with decreasing Re$_\lambda$. This is because the assumption of $\Pi_{\rp} \sim k^0$ is not exactly true, especially for the largest Re$_\lambda$, and only provides a rough estimate of the self-similarity of $E_\rp (k)$. However, the scaling ranges get better with decreasing Re$_\lambda$ also because the fluid non-linearity becomes weaker and elastic effects begin to dominate the flux contributions. This is clearly brought out comparing Fig~\ref{fig:Flux2}(a) and Fig~\ref{fig:Flux2}(b). More importantly, these estimates are obtained by relating the the polymer statistics to that of the fluid velocity field, i.e., the knowledge of one is required to obtain the other. However, the relation between these statistics changes at the smallest Re$_\lambda = 40$ where the fluid nonlinearity remains dormant and falls-off exponentially as shown in Fig.~\ref{fig:Flux2}(c).}

{At the smallest $\Rel \approx 40$, dissipative effects become dominant} so that polymeric contribution $\calp$ to the total flux is of the same order as $\caldf$ for $k \gtrsim 10$ in~\cref{fig:Flux2}(c). Hence, a simple comparison of the fluid dissipation and polymer terms gives (see also~\citet{SinghET23})
\begin{align}
	k^2 \uk \sim k C_k  \implies E_\rp (k) \sim \uk \sim \sqrt{k E_k} \sim k^{-3/2},
\end{align}
which is also consistent with the data shown in~\cref{fig:Epk}(a). 

{We now show the variation of the polymer spectrum with $De$ in~\cref{fig:Epk}(b) for large $\Rel$. At very small $De$, the polymers barely stretch and they quickly relax back to their equilibrium lengths. This means fluctuations in their lengths are largely limited to only small scales and minimal large scale fluctuations. Thus, at relatively large scales and small $k$, for small $De$ polymers, one expects a minimal growth in the spectrum of fluctuations making it appear rather flat. This can otherwise also be understood in terms of real space fluctuations as follows. In real space, as the polymers remain small, their fluctuations are localised to small scales as well such that their fluctuations are correlated over only small scales. So, at large separations, these fluctuations are decorrelated and appear as white noise. This gives the spectrum its flat shape over a wide range of $k$. However, at very large $k$, the spectrum $E_\rp (k)$ falls off much rapidly which is just a consequence of the small scales being well resolved and the polymer fields being analytic. }

{Now, as $De$ increases, polymers begin to stretch to longer lengths, so that their fluctuations are now correlated over longer lengths whose effect begins to be seen in the spectrum being more pronounced at small $k$. At largest $De$, in a close agreement with the scaling arguments~\ref{eq:Ck}, we find that the spectrum $E_\rp (k) \sim k^{-1.35}$ for a wide range. The small scale fall-off of the spectra still confirms the analyticity of the polymer stress field. }

We now move on to real space statistics, by first discussing the {non-unique scaling} nature using structure functions, complementing the discussion in~\cref{sec:Spectra}.  

\subsection{Velocity differences and structure functions}
\label{sec:Sf}

In this section, we discuss how a non-unique scaling behaviour is also manifested in the real-space statistics of PHIT. This complements the Fourier space discussion of~\cref{sec:Spectra}. One of the most common real-space measures that admit a scaling behaviour in turbulence are the well known structure functions $S_\rp(r)$, defined as the $\rp$-th moments of the longitudinal velocity increments $\delta_{\br} u$ over a separation $\br$:
\begin{align}
  \delta_{\br} u  &\equiv \lrs{\ub(\bx + \br) - \ub(\bx)} \cdot \hat{\br},						\label{eq:Ldiff}		\\
   S_\rp(r) = \lra{\lrp{\delta_{\br} u}^\rp} &= \frac{1}{3} \sum_{i=1}^{3} \lra{\lrs{u_i \lrp{x_i + r_i} - u_i\lrp{x_i}}^\rp} 			.			\label{eq:S2def}
\end{align}
\begin{figure}
	\centering
	\includegraphics[width=\textwidth]{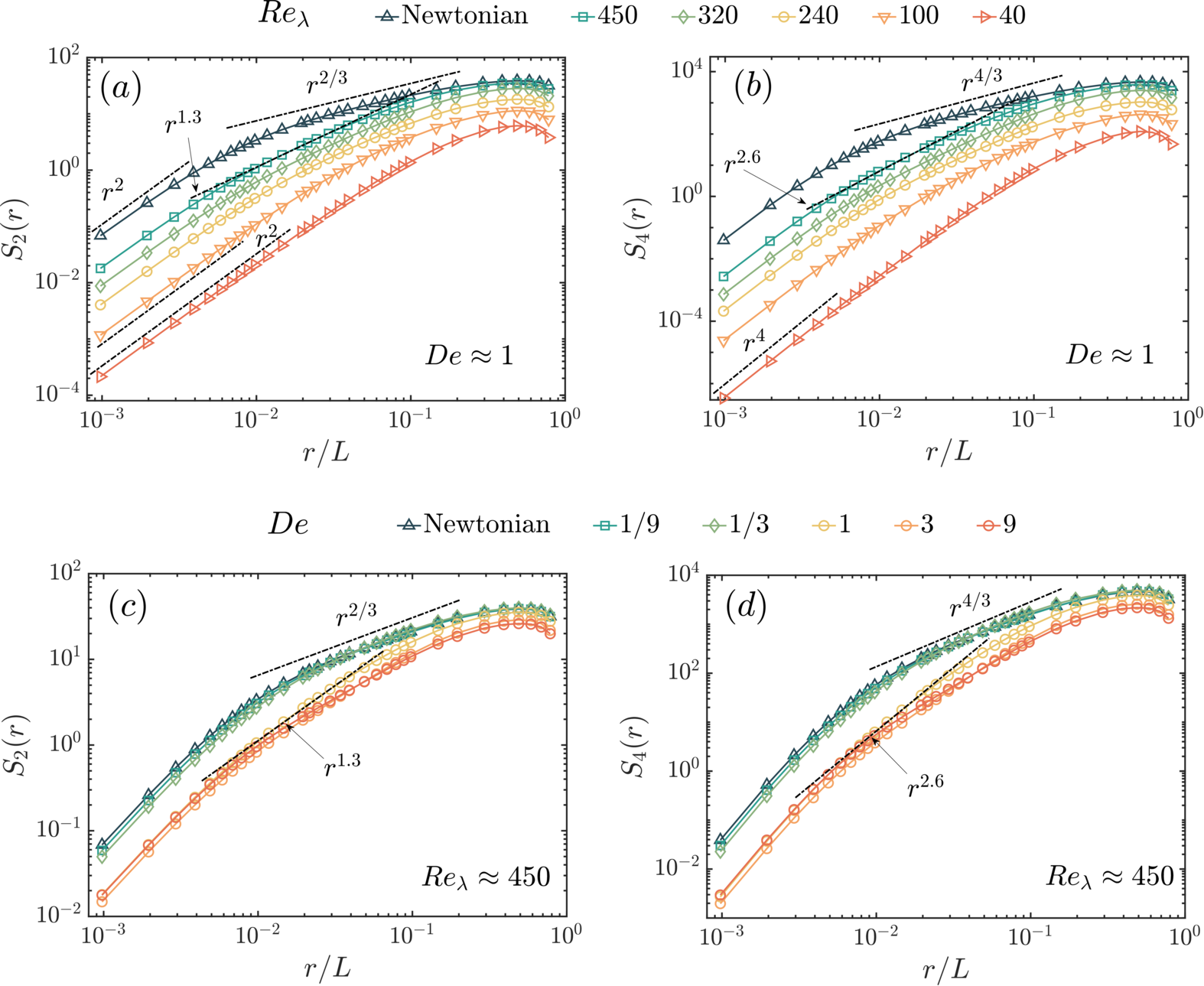}
	\caption{Manifestation of the non-unique scaling behaviour in the structure functions of second ($S_2$: panels a,c) and fourth ($S_4$: panels b,d) orders. (Top) Panels (a,b) show the dependence on $\Rel$, while (Bottom) panels (c,d) show the dependence on the polymer elasticity $De$. Note that the (c) $S_2(r)$ and (d) $S_4(r)$ curves fall on top of the Newtonian for small $De$. At large $De$, the elastic scaling regime becomes clear as elasticity begins to play a significant role showing a clear departure from the Newtonian curve.}
	\label{fig:Sfs}
\end{figure}
In particular, $S_2 (r)$ is related to the energy spectrum $E(k)$ via a Fourier transform. For ease of arguments, we illustrate this using the vector second order structure functions defined  as (see Eq. 6.29, 6.40a in~\citet{Davidson})
\begin{align}
\lra{\lrs{\ub \lrp{\bx + \br} - \ub \lrp{\bx}}^2} \equiv \lra{\lrs{\Delta \ub}^2} = 	\frac{1}{r^2} \frac{\partial }{\partial r}\lrs{ r^3 S_2 (r)} 
\label{eq:S2vect}
\end{align}
The term on the left can be expanded as:
\begin{align}
	\lra{\lrs{\ub \lrp{\bx + \br} - \ub \lrp{\bx}}^2}  &= 2 \lra{\ub^2} - 2 \lra{\ub \lrp{\bx+\br} \cdot \ub \lrp{\bx} } \nonumber \\
	&= 4 \int d \bk \  |\uh(\bk)|^2 - 4  \int d \bk  |\uh(\bk)|^2   e^{-i \bk \cdot \br}  \nonumber	\\
	& = 4 \int dk \ E(k) \lrp{1-\frac{\sin{kr}}{kr} } \label{eq:S2Ek}
\end{align}
where we have the definition of Fourier transforms to obtain the second line and isotropy to obtain the last line. The previous relation can then be easily used to relate the power-law behaviour of $S_2$ and $E(k)$. Consider that the separation $r$ in the real-space is scaled by a factor $b$, i.e. $r \to b r$, so that the Fourier space wave number scales as $k \to b^{-1}k$. Consequently, if the spectrum scales with an exponent $\alpha$, i.e. $E(k) \sim k^{-\alpha}$, then $E(k) \to E(b^{-1}k) = b^{\alpha} E(k)$. Therefore, the right hand side of~\cref{eq:S2Ek} can be rewritten under this rescaling as: 
\begin{align}
\lra{\lrs{\Delta \ub}^2}_{\textit{rescaled}} &=	4 \int \lrp{b^{-1} dk}\  E(b^{-1}k) \lrp{1-\frac{\sin{kr}}{kr} } 	\nonumber\\
	&= 4 b^{\alpha-1} \int dk \ E(k) \lrp{1-\frac{\sin{kr}}{kr} } = b^{\alpha-1} \lra{\lrs{\Delta \ub}^2}.
\end{align}
Thus, it is easy to see that the vector and longitudinal structure functions transform identically under rescaling since
\begin{align}
& \lra{\lrs{\Delta \ub}^2}_{\textit{rescaled}} =	\frac{1}{ (br)^2} \frac{\partial }{\partial (br)}\lrs{ (br)^3 S_2 (br)} = \frac{1}{ r^2} \frac{\partial }{\partial r}\lrs{ r^3 S_2 (br)}  	\nonumber \\
& =  b^{\alpha-1} \lra{\lrs{\Delta \ub}^2} = \frac{1}{r^2} \frac{\partial }{\partial r}\lrs{ r^3 b^{\alpha-1} S_2 (r)} 	\implies S_2 (br) = b^{\alpha-1} S_2 (r).
\end{align}

Hence, we have that $E(k) \sim k^{-\alpha} \iff S_2(r) \sim r^{\alpha-1}$. Note that this relation holds only for $\alpha \leq 3$. For $\alpha > 3$, the real-space exponent is constrained by the leading order in Taylor expansion as $S_2 \sim r^{2}$. In HIT, these arguments imply that $S_2 \sim r^{2/3}$ given $E(k) \sim k^{-5/3}$ whereas in PHIT we now expect $S_2 \sim r^{1.3}$ using $E(k) \sim k^{-2.3}$. Additionally, using dimensional arguments, one also expects within the K41 phenomenology that $S_4 \sim S_2^2 $. With these expected results in mind, we show in~\cref{fig:Sfs}(a,b) the plots of $S_2(r), S_4(r)$ for PHIT at different $\Rel$ and $De \approx 1$ alongside the HIT result. The variation of $S_2(r)$ and $S_4(r)$ with $De$ is shown in~\cref{fig:Sfs}(c,d). We of course have that the HIT structure functions follow the Kolmogorov scaling $r^{\rp/3}$ closely in~\cref{fig:Sfs}(a,b), while those for PHIT show a very clear departure from this behaviour. Indeed, and in consistency with the discussion above, the \poly scaling is given by $S_2 \sim r^{1.3} (S_4 \sim r^{2.6})$ and is seen to have the widest span at the largest $\Rel \approx 450$. This is also consistent with the discussion in~\cref{sec:Spectra} where we show that the \poly regime shrinks with decreasing $\Rel$, while the smooth dissipation range marked by $S_2 \sim r^2$ expands (corresponding to $E(k) \sim k^{-4}$ in~\cref{Ek}). At the smallest $\Rel \approx 40$, the smooth scaling $S_2 \sim r^2$ spans almost a decade. 

We show the variation of the structure functions with $De$ in~\cref{fig:Sfs}(c,d). As expected from~\cref{Ek}(a), this dependence is rather non-monotonic. The small $De$ cases show an approximately Newtonian behaviour, owing to the minimal non-Newtonian contributions as the polymers strongly resist any stretching (see~\cref{sec:Spectra} for detailed discussion). As $De \to 1$, we have a clear \poly scaling $S_2 \sim r^{1.3} (S_4 \sim r^{2.6})$. This steeper than Kolmogorov slope starts to vanish when polymer elasticity is further increased, and the $S_2/ S_4$ curves now have a shallower slope. At much larger polymer elasticity one indeed recovers the Kolmogorov scaling (see~\citet{Marco23}). This behaviour is again in consistency with that for the spectrum in~\cref{sec:Spectra}.

We also make an important note about~\cref{fig:Sfs}(b,d). The $S_4(r)$ curves evidently show smaller scaling ranges compared to those of $S_2(r)$ in~\cref{fig:Sfs}(a,c). This means that the exponents quickly deviate from the naive expectation of $S_4(r) \sim S_2^2(r)$. This is not only an artefact of the deviations from a K41 like behaviour but also captures the non-Gaussianity of $\delta_{\br} u$. We quantify these deviations by measuring the \textit{kurtosis} of the velocity increment distributions in the next section.

\subsection{Kurtosis of velocity differences}
 \label{sec:Kurt}
 \begin{figure*}
 	\centering
 	\includegraphics[width=\textwidth]{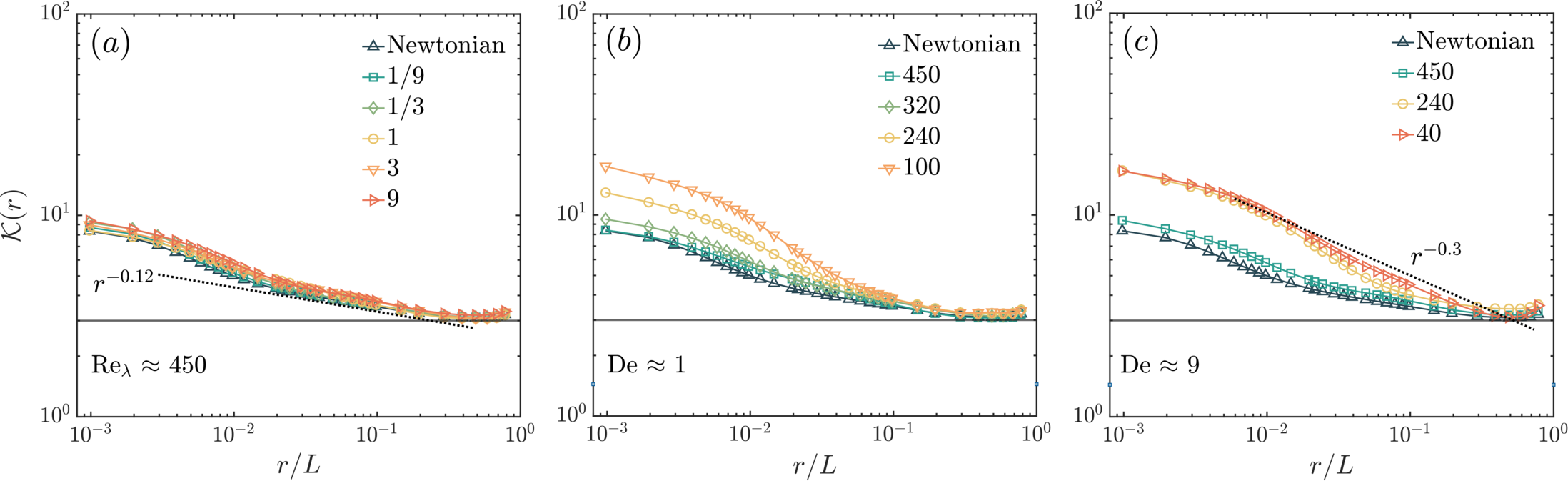}
 	\caption{The kurtosis of velocity differences as a function of scale $r$. $\kurt \to 3$ as $r \to L$ as velocity differences decorrelate at very large separations. (a) $\kurt$ at large $\Rel \approx 450$ and different $De$ are almost coincident implying a very weak dependence of intermittency on $De$.  (b,c) Intermittency in polymeric flows increases as $\Rel$ is decreased, especially when $De$ is large.}
 	\label{fig:Kurt}
 \end{figure*} 

Consider the distribution of (longitudinal) velocity differences $\delta_{\br} u$, defined by~\cref{eq:Ldiff}, as a function of the scale $r$. The kurtosis (or flatness) $\mathcal{K}(r)$ of such a distribution is defined as:
\begin{align}
	\kurt (r) \equiv \frac{\lra{\lrp{\delta_{\br} u}^4}}{\lra{\lrp{\delta_{\br} u}^2}^2}.				\label{eq:Kurt}
\end{align}
The kurtosis $\kurt$ captures the relative importance of the tails of a distribution and quantifies their contribution to the overall statistics of $\delta_{\br} u$. For a Gaussian distribution, $\kurt = 3$. We expect the velocity increments to be uncorrelated over very large separations, so that the distribution of $\delta_{\br} u$ is close to a Gaussian for large $r$. This is indeed confirmed in our data in~\cref{fig:Kurt} where we find $\mathcal{K}(r) \approx 3$ for large $r$, irrespective of $\Rel$ and $De$. More importantly, $\kurt$ is not a constant but a function of the scale $r$.  The distributions of $\delta_{\br} u$ are, evidently, devoid of scale-invariance and their departures from Gaussianity become increasingly stronger at small $r$. This is a result of more important contributions from the tails of the distribution of $\delta_{\br} u$, a phenomenon referred to as intermittency, and causes deviations from the dimensional expectation of $S_4 \sim S_2^2$ (this would have meant $\kurt \sim r^0$ from~\cref{eq:Kurt}). The log-log plots of $\kurt$ in~\cref{fig:Kurt}(a) clearly show a power-law dependence on $r$ at large $\Rel \approx 450$. Crucially, this power-law is very weakly dependent on $De$ as the curves for different $De$ follow each-other closely. Thus, the intermittent nature of the velocity fluctuations is still largely determined by the fluid non-linearity. This was also observed by~\citet{Marco23} via the multifractal spectrum for energy dissipation. The influence of polymers on velocity fluctuations, however, is more apparent at small $\Rel$, as shown in~\cref{fig:Kurt}(b). At small $\Rel$, although the fluid non-linearity becomes considerably weaker (see~\cref{fig:Flux2}(b)), the polymer stresses feeding back on the flow result in the velocity fluctuations $\delta_{\br} u$ being more intermittent, and the departure from HIT curve becomes significant. This is even more pronounced at the smallest $\Rel$ (we considered) where the presence of polymers results in much large deviations of $\delta_{\br} u$ at small $r$ which is manifested in a much steeper slope of the kurtosis $\kurt (r) \sim r^{-0.3}$ compared to the Newtonian. 

So, the departure from a Kolmogorov-like behaviour in polymeric flows is not only due to a modified energy flux---resulting in a different scaling of the spectrum---but also due to the modified nature of extreme fluctuations that result from the action of the polymer stresses. This also underlines the fact that intermittency is due not only to fluid non-linearity but even polymer stresses can drive highly intermittent velocity fluctuations. Such intermittent behaviour is also discernible in the statistics of velocity gradients. In the following sections, we discuss the intermittent nature of velocity gradients, and how they influence the stretching of polymers. This also forms the basis for the discussion of the two energy dissipation rates: fluid $\epsilon_{\rf}$ and polymer $\epsilon_{\rp}$.

\subsection{Polymer stretching, velocity gradients, and their relation to the energy dissipation}
\label{sec:Trace}

\begin{figure*}
	\centering
	\includegraphics[width=\textwidth]{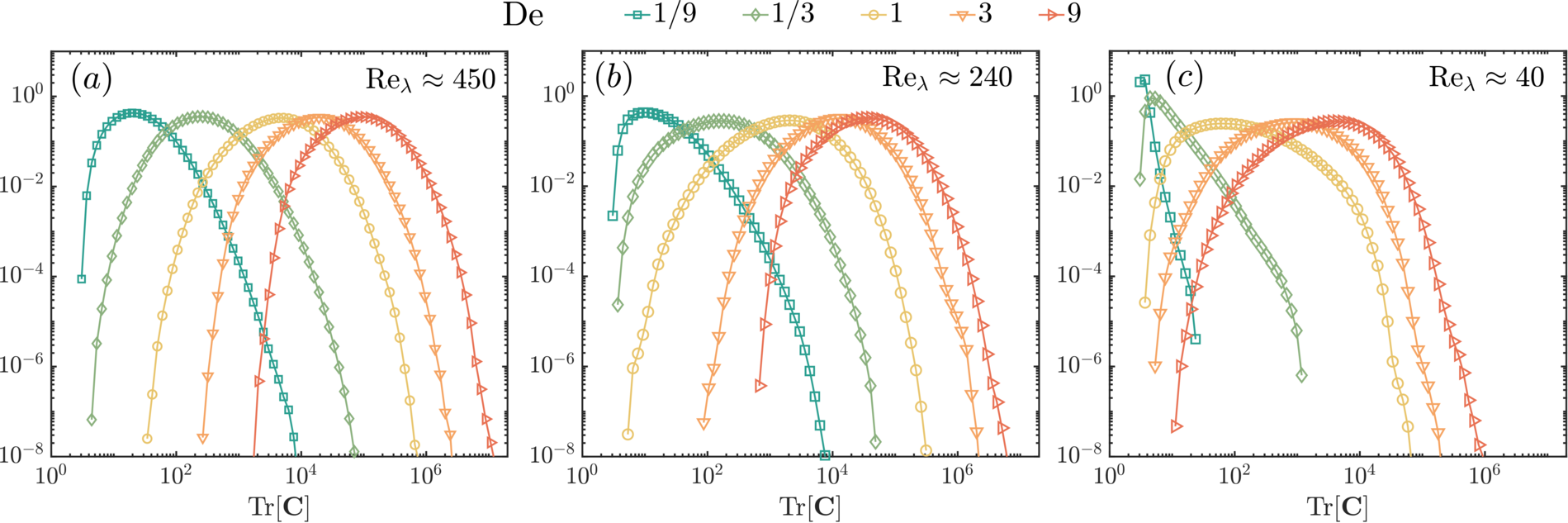}
	\caption{Pdfs of the end-to-end polymer lengths $\trC$ for different $\Rel$: (a) $\Rel \approx 450$, (b) $\Rel \approx 240$, and (c) $\Rel \approx 40$ and all $De$. A large $De$ polymer has a larger average end-to-end length, as seen in the right shift of the pdfs.}
	\label{fig:TrC1}
\end{figure*} 
We now discuss in this section how polymers of different $De$ are themselves affected by the fluid flow at different $\Rel$. To this end, we begin by simply looking at the stretching statistics of the polymers. measure of the stretching of the polymers is their end-to-end lengths whose instantaneous squared value is given by $\trC$. We naturally expect polymers to stretch more with increasing $De$ at any given $\Rel$. {As stated earlier, in the limit of $De \to 0$ the polymers are almost inextensible and $\C$ doesn't stray far away from its equilibrium configuration $\I$. On the contrary, the other limit of $De \to \infty$ implies a weak restoring force, allowing $\C$ to grow to large values. That is, polymers take a very long time to relax back to their equilibrium lengths for a very large $\tau_{\rm p}$. Or in other words their tendency to relax back, when extended, is much lesser.} Evidently, polymers are stretched by the local velocity gradients given by the first two terms on the r.h.s of~\cref{Conf}.
\begin{figure*}
	\centering
	\includegraphics[width=0.8\textwidth]{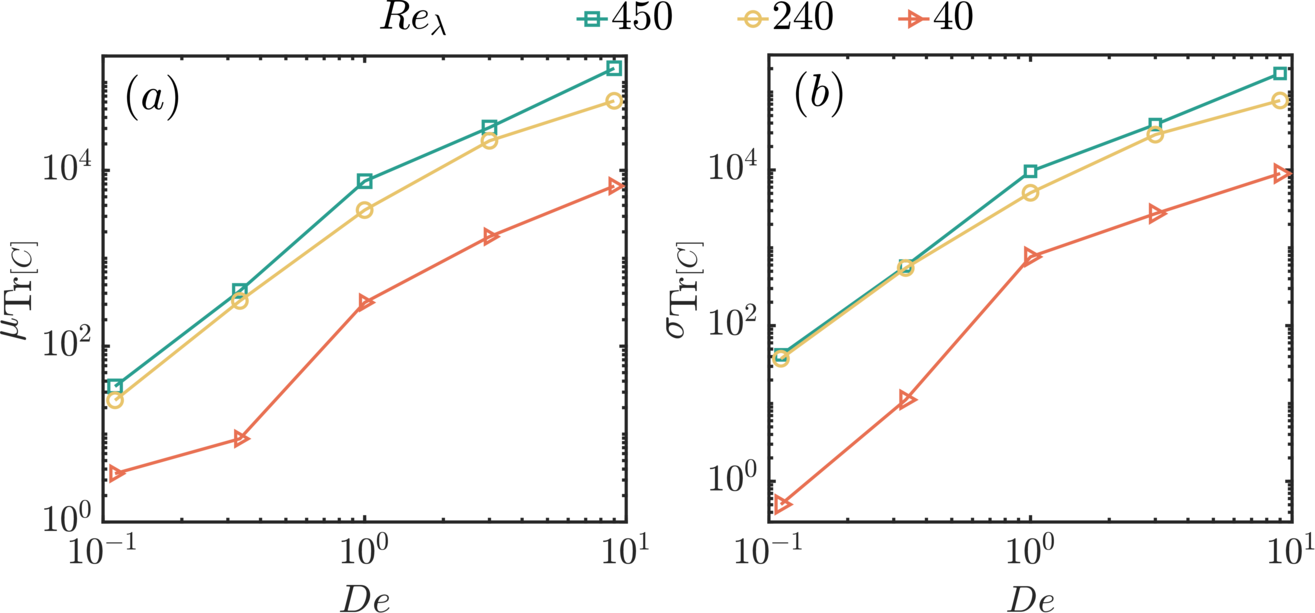}
	\caption{(a) Mean $\mu_\trC$ and (b) standard deviation $\sigma_\trC$ of the distributions in~\cref{fig:TrC1}, shown as a function of $\Rel$ and $De$. }
	\label{fig:TrC2}
\end{figure*} 

We plot the pdfs of $\trC$ for different $\Rel$ in~\cref{fig:TrC1}. We first note that although polymers described by the Oldroyd-B model are infinitely stretchable, they can only do so in the presence of very large and sustained velocity gradients. This is confirmed by the very fast decay of the tails of the $\trC$-pdfs in~\cref{fig:TrC1}, clearly suggesting that the polymer lengths always remain bounded in our simulations. \cref{fig:TrC1}(a) shows that, at large $\Rel$, the pdfs shift towards the right with increasing $De$ indicating a larger averaged polymer length. Moreover, the width of the distributions also increases with $De$, meaning that polymer lengths also fluctuate over a wider range. This is more accurately shown by~\cref{fig:TrC2} where we plot the mean ($\mu_{\trC}$) and standard deviation ($\sigma_{\trC}$) of the distributions as a function of $\Rel$ and $De$. Indeed, the polymer lengths and their fluctuations increase with both $\Rel$ and $De$. 

Panels (b) and (c) in~\cref{fig:TrC1} show the $\trC$ distributions for smaller $\Rel$. With decreasing $\Rel$, turbulence intensity decreases and the velocity gradients become weaker on an average and stretch the polymers less (as can be seen from~\cref{Conf}). Therefore, the mean polymer lengths decrease with decreasing $\Rel$ as the pdfs shift towards smaller values of $\trC$. Moreover, polymers are also limited in their maximum extensions as $\Rel$ decreases. At the smallest $\Rel$, the velocity gradients are unable to stretch the less extensible, small $De$ polymers considerably. This is captured in~\cref{fig:TrC1}(c), where the $\trC$-pdfs fall-off very fast at small $De$ showing that polymer lengths have rather small fluctuations. (This also implies that polymer stresses remain very small so that the flow remains laminar for $De < 1$). However, for a large enough $De (\approx 1)$, polymers stretch to considerably long lengths, and the resulting large fluctuations are enough to excite sub-dominant velocity fluctuations which gives rise to ET with $E(k) \sim k^{-4}$ (see~\cref{Ek}). 

\begin{figure*}
	\centering
	\includegraphics[width=1.0\textwidth]{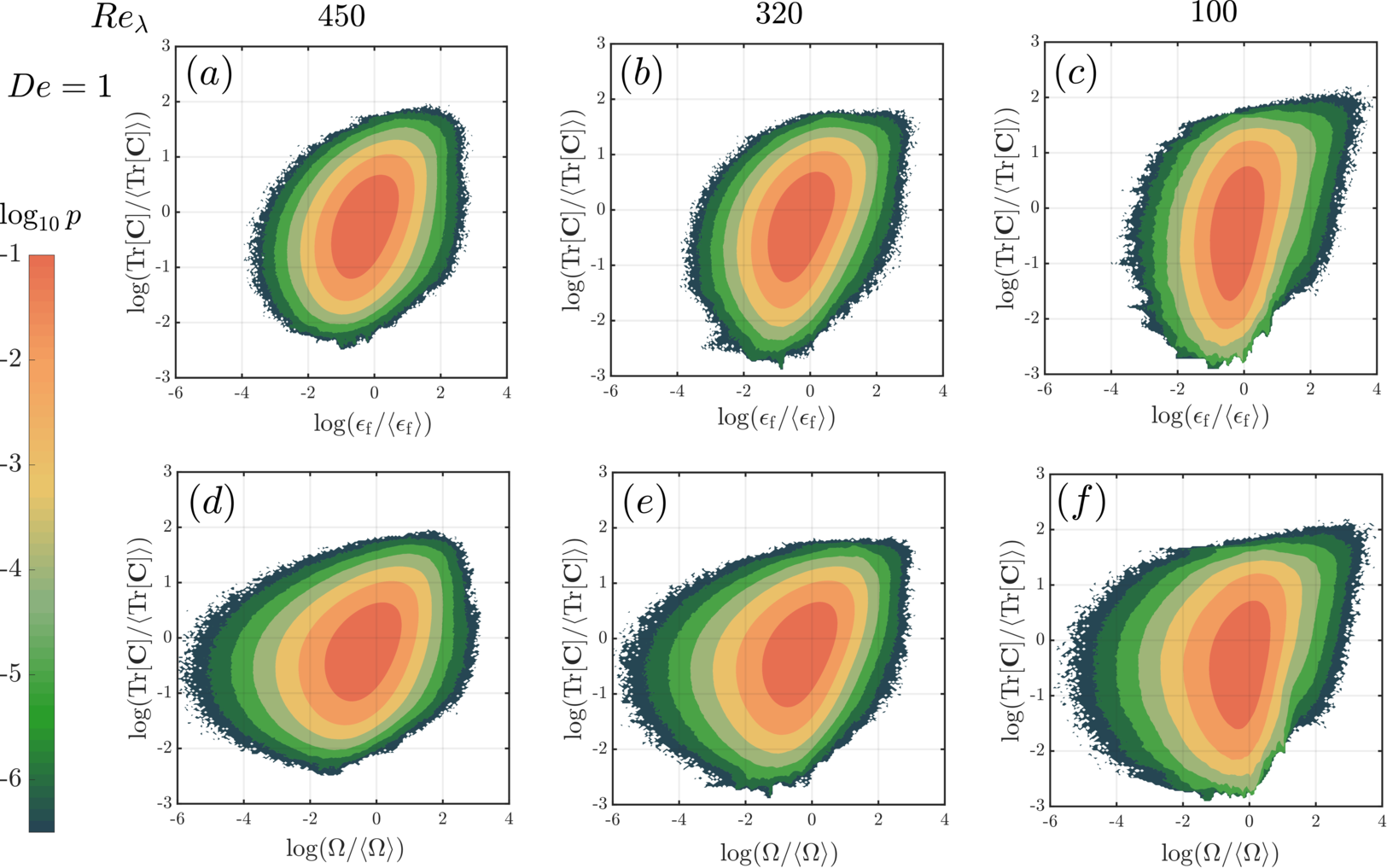}
	\caption{The plot of the joint distributions of the polymer lengths on log-log scale with the (top) dissipation rate $\epsilon_{\rf}$ and (bottom) enstrophy $\Omega$, for different $\Rel$ with $De \approx 1$. The color bar shows the color coding of the logarithm of probability $\log_{10} p$.}
	\label{fig:Jpdf1}
\end{figure*} 
The above discussion is largely based on how polymers with different $De$ are stretched by carrier flows at different $\Rel$. To understand this better, we now discuss how the polymer stretching $\trC$ is correlated with the velocity gradients $\Aij (\bx,t) \equiv \pari u_j (\bx,t)$ in the flow. To this end, we first decompose $\Aij$ into its symmetric $S_{ij}$ and antisymmetric $\omega_{ij}$ components that respectively capture the purely extensional (compressional) and the purely rotational contributions. The squared magnitude of these components are given by the fluid dissipation rate $\epsilon_{\rf}$ and enstrophy $\Omega$:
\begin{align}
	\Omega = \omega_{ij} \omega_{ij} &\ ; \quad 	\epsilon_{\rf} = 2 \nu S_{ij}S_{ij},		\nnn
	\text{where} \quad  \omega_{ij} = \lrp{\pari u_j - \parj u_i  }/2  &\ ; \quad    S_{ij} = \lrp{\pari u_j + \parj u_i }/2.
\end{align}
The presence of the symmetric term $\C \nabla \ub + (\nabla \ub)^T \C = 2 C_{ik} S_{kj}$ in~\cref{Conf} means that the extensional (compressional) regions of the flow are directly responsible for the stretching (relaxation) of the polymers. On the other hand, the absence of any $\omega_{ij}$ terms means that there is no direct effect of the rotation in changing the lengths of polymers. However, turbulence in three-dimensions is marked by the stretching and intensification of vorticity in the direction of local velocity gradients. So, the vortical regions can be expected to correlate with large polymer lengths, although to a lesser degree than the straining regions. (A direct computation of Pearson correlation coefficients of the joint distributions of $(\trC,\epsilon_{\rf})$ as well as those of $(\trC,\Omega)$ yielded values $\approx 0.3$ for all cases, while those of the $\log$ of normalised variables which are actually studied, e.g. of $(\log [\trC/\lra{\trC}], \log[ \epsilon_{\rf}/\lra{\epsilon_{\rf}}])$, yielded values $\approx 0.4$ showing that indeed these variables have a non-zero, finite correlation.)

To illustrate the above, we show in~\cref{fig:Jpdf1} the joint probability distributions (jpdfs) of the polymer lengths (given by $\trC$) with the local stretching rates (given by $\epsilon_{\rf}$) as a function of $\Rel$ for one instance of polymer elasticity of $De = 1$ in the top panel. We observe that the longest end-to-end lengths of the polymers indeed coincide with the largest stretching rates, while the smallest polymers are found in regions of the flow with small extensional rates. This results in a perceivable tilt of the joint distribution in panel (a). With decreasing $\Rel$, the mean lengths of the polymers decreases (see~\cref{fig:TrC2}) and their preference for regions with small scale fluctuations increases. This is evident from~\cref{fig:Jpdf1}(b,c) where elongated bottom tails of the jpdfs imply small polymers lengths are more probable at smaller $\Rel$. While we still have that the largest polymer extensions coincide with the largest velocity gradients, the core of the jpdfs is now more vertical which indicates that much larger extensions are now possible at even the average stretching rates. This hints towards a decreasing degree of correlation between extreme stretching rates and polymer stretching as $\Rel$ is reduced.

\begin{figure*}
	\centering
	\includegraphics[width=1.0\textwidth]{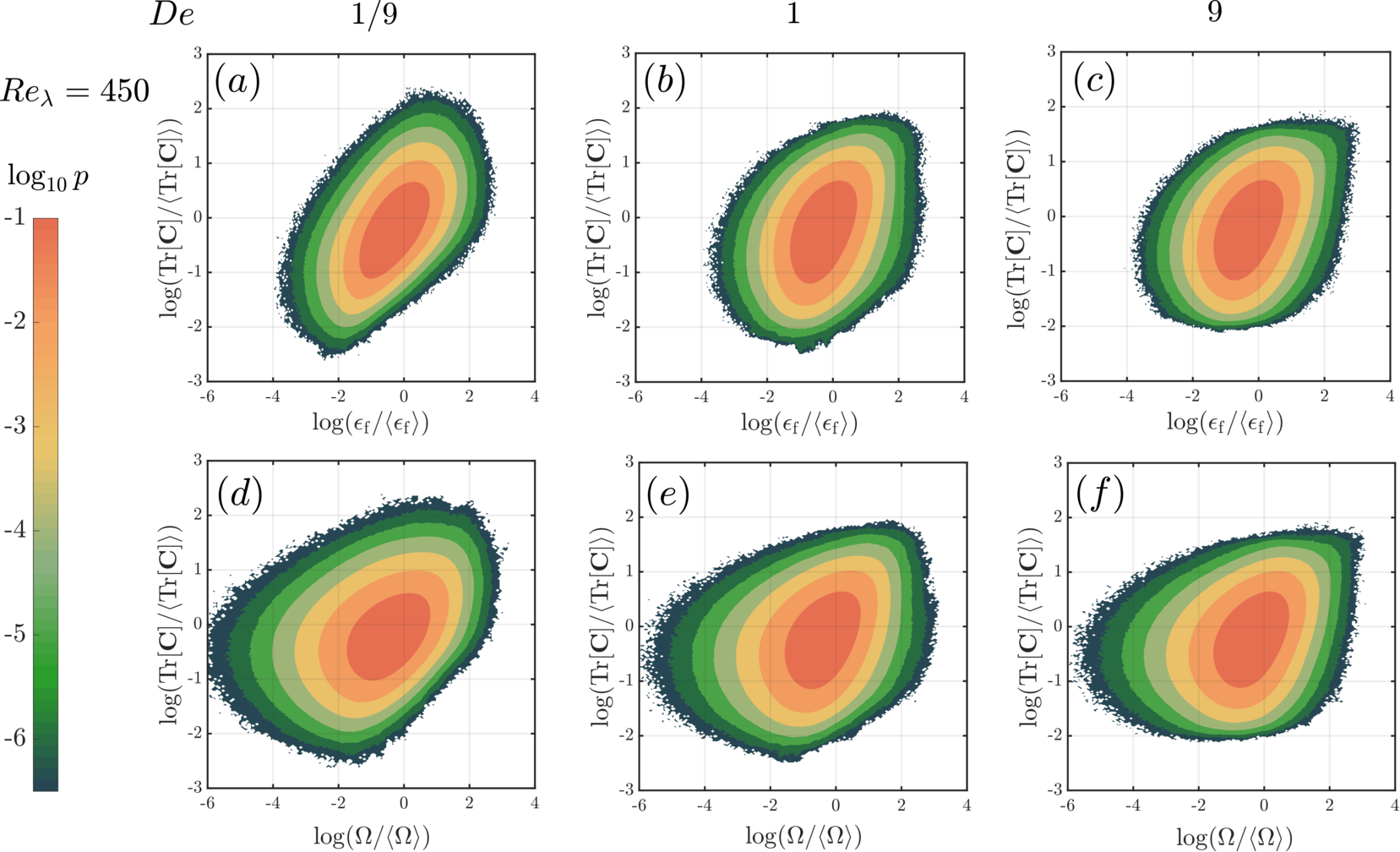}
	\caption{Joint distributions of the polymer lengths with the (top) $\epsilon_{\rf}$ and (bottom) $\Omega$, for different $De$ at $\Rel \approx 450$. Different colors correspond to the logarithm of probability $\log_{10} p$ as coded by the colorbar. }
	\label{fig:Jpdf2}
\end{figure*} 
The bottom panel of \cref{fig:Jpdf1} shows the jpdfs of $\trC$ and enstrophy $\Omega$. At large $\Rel$, large polymer lengths also coincide with extreme vorticity regions. This means polymers are also stretched in regions of large vorticity (similar to what observed for long, polymer-like fibres in~\citet{Jason2020}). The maximum enstrophy regions, however, are more likely to have polymers with close to average lengths, implied by~\cref{fig:Jpdf1}(d), unlike the straining regions whose extremes see maximally extended polymers. More importantly, polymers are  likely to be stretched even in regions with very small $\Omega$ as suggested by the left bulge of the jpdfs. This shows that the polymer stretching-enstrophy correlation is not as strong as the stretching-straining correlation. This is even more evident with decreasing $\Rel$ as vortex stretching becomes progressively weaker. Indeed, panel (f) shows that polymer lengths are rather uniformly distributed, especially for small enstrophy, indicating an even less correlation with polymer stretching.

\cref{fig:Jpdf2} shows instead how polymer stretching correlates with the local stretching (top panel) and rotation rates (bottom panel) of the flow, as a function of polymer elasticity. It was already shown that large $De$ polymers stretch more and is clearly captured by their mean lengths in~\cref{fig:TrC2}. {At small $De$, polymers stretch minimally and have small relaxation times.} Therefore, while they are stretched by regions with very large stretching rates, those regions have rather small lifetimes (a quick estimate gives $\tau \sim S_{ij}^{-1}$). So, maximally stretched polymers are more likely to persist for longer in regions with moderate stretching rates. This means that the maximum of $\trC$ coincides with only slightly larger than the average values of $\lra{\epsilon_{\rf}}$. However, at larger $De$, polymers are more stretchable and easily acquire their maximal lengths upon encountering strong straining regions (similar to the observation in~\citet{Singh2024} that long, elastic, polymer-like fibres preferentially sample straining regions). At these large $De$, polymers remain stretched for longer times so that the fluctuations about the mean become smaller, across all values of $\epsilon_{\rf}$, thus giving the jpdfs more rounded contours. 
The bottom panel of~\cref{fig:Jpdf2} plots the $\trC-\Omega$ joint distributions. We, of course, again have a much wider spread of the distributions compared to $\trC-\epsilon_{\rf}$, yet again suggesting that enstrophy and polymer stretching are less correlated compared to $\epsilon_{\rf}-\trC$, as polymers can also be stretched when enstrophy is very small. Thus, while the maximally stretched polymers are most likely found in regions with maximal enstrophy at large $De$, they also tend to be stretched in regions with very small $\Omega$ when $De$ is large.

\subsection{Energy dissipation}
\label{sect:Diss}
\begin{figure*}
	\centering
	\includegraphics[width=1.0\textwidth]{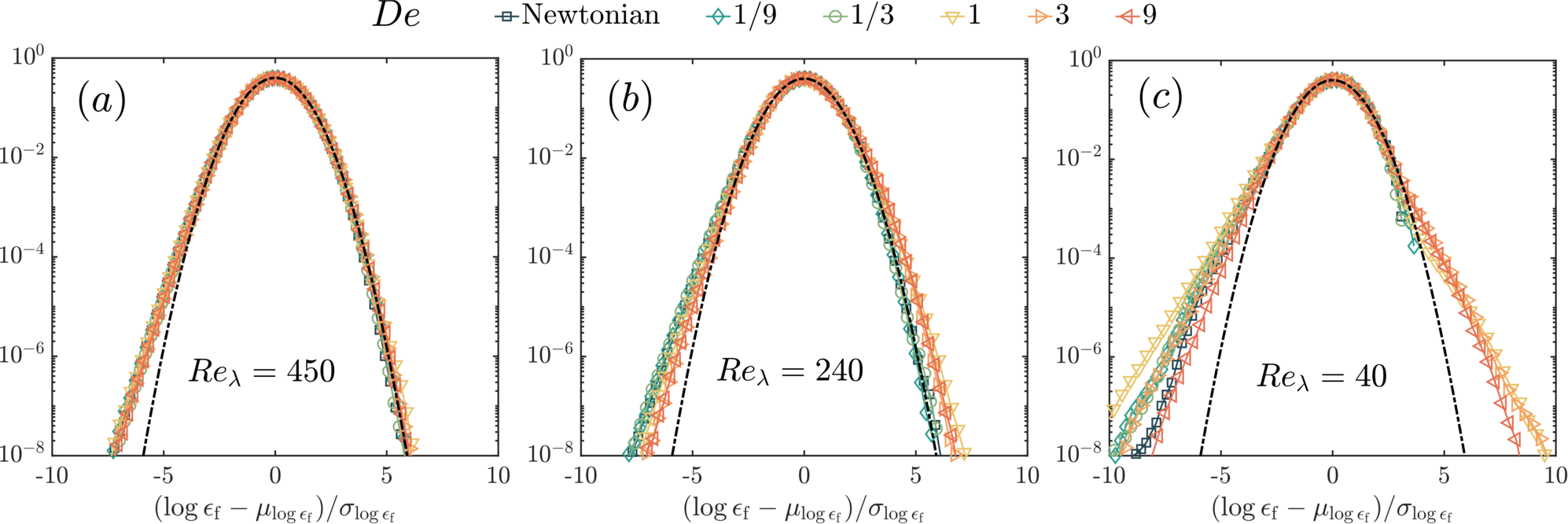}
	\caption{The pdf of the fluid energy dissipation rate $\epsilon_{\rf}$, compared to a log-normal distribution, for different $De$ and $\Rel$. At (a) $\Rel \approx 450$, polymer elasticity has no influence on the intermittent nature of the distributions. However at smaller (b) $\Rel \approx 240$, the presence of large $De$ polymers leads to more large deviations, which are even more prominent at (c) $\Rel \approx 40$.}
	\label{fig:Dissf}
\end{figure*} 
\begin{figure}
	\centering
	\includegraphics[width=\textwidth]{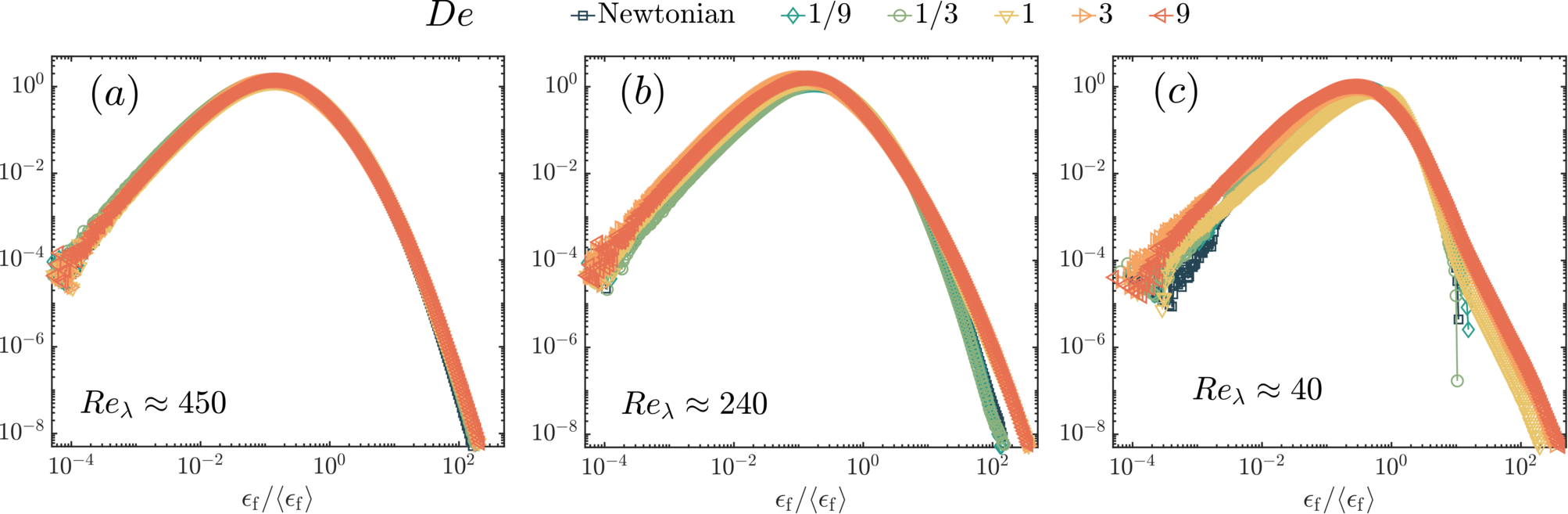}
	\caption{The pdf of the fluid energy dissipation rate $\epsilon_{\rf}$ for different $De$ and $\Rel$. Each panel corresponds to a decreasing $\Rel$ from left to right, for all the investigated $De$.}
	\label{fig:Epsf}
\end{figure}
\begin{figure}
	\centering
	\includegraphics[width=0.45\textwidth]{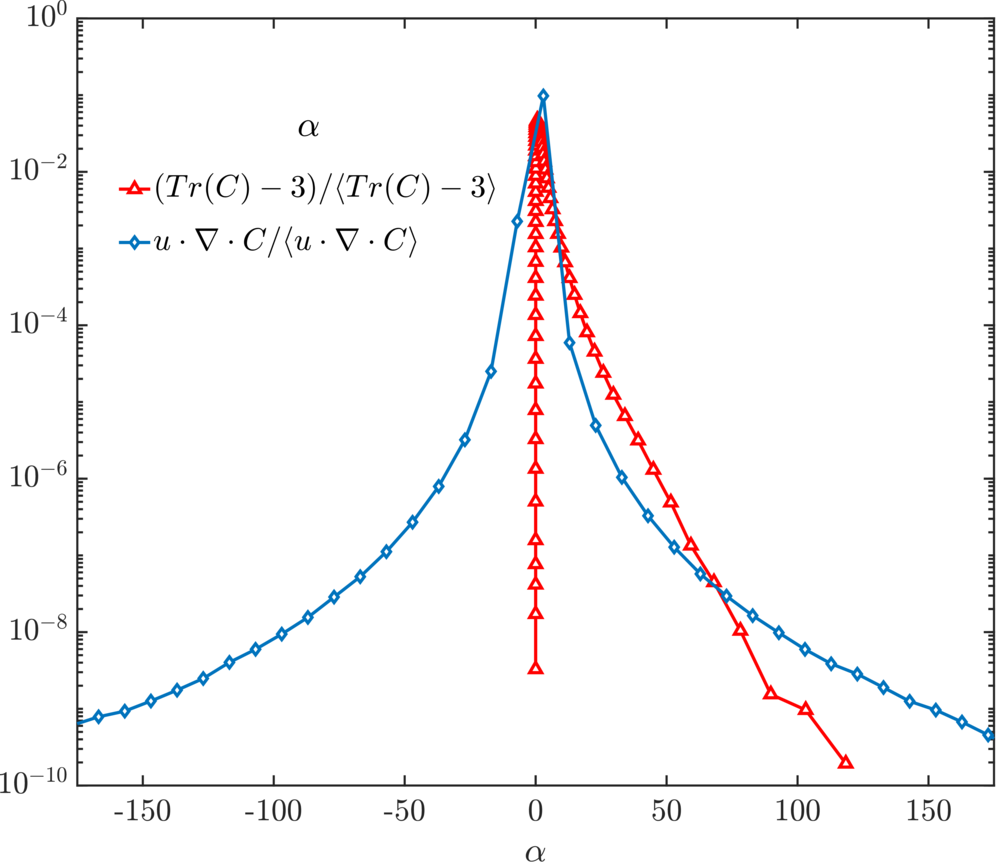}
	\caption{The pdfs comparing the local dissipation and transfer processes associated with polymers. They equal each other only on an average, while locally the processes are very different. }
	\label{fig:comp}
\end{figure}

In this last section, we discuss how the nature of energy dissipation in turbulence is modified in the presence of polymers. In polymeric flows, energy is dissipated away by both the carrier fluid $\epsilon_{\rf}$ and the dissolved polymers $\epsilon_{\rp}$. The positive definite fluid dissipation is given by $\epsilon_{\rf} = 2\nu S_{ij}S_{ij}$, whereas the positive definite polymeric dissipation is related to the average end-to-end polymer length by $\epsilon_{\rp} = (\mu_\rp/2\taup^2) \lra{\trC -3}$. Both these relations can be readily obtained by multiplying~\cref{NN} with $\ub$, and using~\cref{Conf} as follows:
\begin{align}
&	\pt \lra{\ub^2}/2 = \nu \lra{\ub \nabs \ub} + \nup \lra{\adv \cdot \C} = - \nu \lra{\lrp{ \nabla \ub}^2} - \nup \lra{\text{Tr} \lrs{\C \nabla \ub}} + \lra{\mathbf{F} \cdot \ub}				\nnn
\implies  &	0 = -2 \nu \lra{S_{ij}S_{ij}} - \nup \lra{\text{Tr} \lrs{\C \nabla \ub}} + \lra{\mathbf{F} \cdot \ub} = - \epsilon_{\rf} - \epsilon_{\rp} + \epsilon_{\rm I}.
\end{align}
Similarly,
\begin{align}
&		\mu_\rp \pt \lra{\C} = \mu_\rp \lra{\C \lrp{\nabla \ub}} + \mu_\rp \lra{\lrp{\nabla \ub}^T \C} - \nup \lra{\C - \I}		\nnn
\implies  &		0 =   \frac{2 \mu_\rp}{ \taup}\lra{\text{Tr} \lrs{\C \nabla \ub}} -  \nup \lra{\trC - 3} 		\implies		\epsilon_{\rp} = \frac{\mu_\rp}{2 \taup^2} \lra{\trC - 3},
\end{align}
where we have used stationarity, periodicity and incompressibility to set time derivatives, fluid non-linearity, pressure contribution and polymer advection terms to zero. Note that, the previous relation $\epsilon_{\rp} =  \nup \lra{\text{Tr} \lrs{\C \nabla \ub}} = \frac{\mu_\rp}{2 \taup^2} \lra{\trC - 3}$ holds true only on average, and thus the two processes are equal only in a statistically stationary state, but very different instantaneously. The $ \text{Tr} \lrs{\C \nabla \ub}$ term captures the exchange of energy between fluid and the polymers, and can take both positive and negative values, while $\lra{\trC - 3}$ is a positive definite quantity and captures purely dissipative effects. The interaction term $ \nup\text{Tr} \lrs{\C \nabla \ub}$ leads to a net transfer of energy to polymers, which is dissipated away as $\frac{\mu_\rp}{2 \taup^2} \lra{\trC - 3}$. This is seen from~\cref{fig:comp}, where we plot the normalized distributions of the two terms using data from our simulations.
 The positive-definiteness of $\epsilon_{\rf}$ and $\epsilon_{\rp}$ means that the total energy injected by the forcing $\epsilon_{\rm I}$ is dissipated away by both polymers and the fluid in a stationary state. The internal energy of the polymers is proportional to their instantaneous extension so that elongated polymers dissipate away energy as they relax. We now discuss the nature of these measures of dissipation in our polymeric flows.
\begin{figure*}
	\centering
	\includegraphics[width=1.0\textwidth]{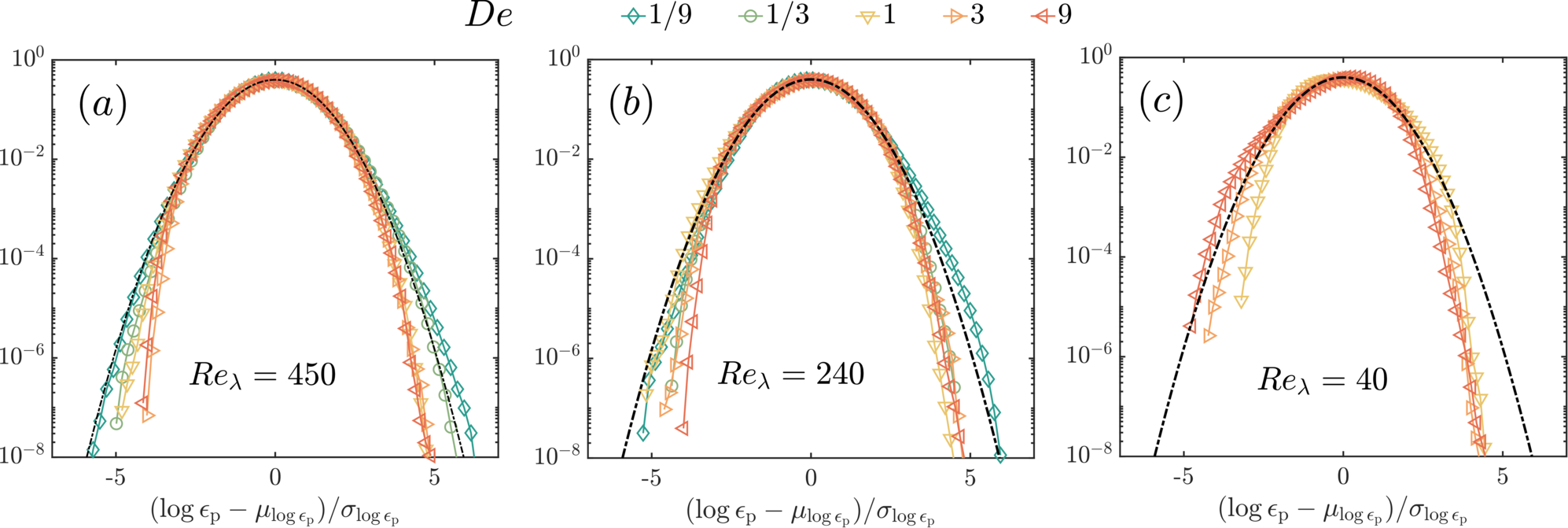}
	\caption{The pdf of the logarithm of polymer dissipation $\epsilon_\rp \propto \trC$, as a function of $De$ for different $\Rel$ given in the panels. The black dashed curve shows a log-normal distribution for reference. The polymer dissipation becomes less intermittent with increasing $De$.}
	\label{fig:Dissp} 
\end{figure*} 

We first recall that $\epsilon_{\rf}$ in HIT has a distribution whose tails deviate from a log-normal behaviour, in contradiction to the~\citet{K62} prediction. We now show in~\cref{fig:Dissf}(a) that the normalised $\log( \epsilon_{\rf})$-pdfs remain largely unaffected by the addition of polymers in flows at large $\Rel$. The curves for different $De$ all collapse over the Newtonian curve in~\cref{fig:Dissf}(a) which shows that spatio-temporal fluctuations in energy dissipation are unaffected by the addition of polymers. Polymer stresses, though present, do not result in yet larger deviations which are still dominated by the fluid non-linearity. (Note that the mean fluid dissipation $\lra{\log\epsilon_{\rf}}$ is still a function of $De$.) At a smaller $\Rel \approx 240$, the Newtonian curve shows an even larger deviation from log-normality, especially at the left tails, suggesting that dissipation field now has relatively more quiescent regions. There is no appreciable change in the distributions upon the addition of small $De$ polymers as the polymer stresses remain small. However, when $De$ is large, elastic effects become appreciable compared to the weakened fluid non-linearity. The polymer stresses are now large enough to create significant fluctuations in the flow so that now there are less quiescent regions and more extreme events. This is apparent from the shrinking left tails and the widening right tails at $De = 1,3$ in~\cref{fig:Dissf}(b). Curiously, at the largest $De = 9$, when the elastic effects begin to weaken again, the large deviations are reduced (this is in parallel to the weakened elastic effects resulting in the shrinking \poly range in~\cref{Ek}(b)). These effects are most clearly seen at the smallest $\Rel = 40$ where fluid non-linearity does not play a role (and decays exponentially in scales, see~\cref{fig:Flux2}(c)). The fluctuations in velocity gradients, therefore, only result from the strong polymer stresses at large $De$ and are manifested in the wide-tailed, super log-normal distributions of $\epsilon_{\rp}$ in~\cref{fig:Dissf}(c). However, for a large polymer elasticity such as $De = 9$, a weakening of elastic effects results in a mild shrinking of the tails of the pdf. 
\cref{fig:Epsf} shows the normalized pdfs of the bare $\epsilon_{\rf}$ for 3 different Re$_\lambda$ and all $De$. It is clear from these plots that at large Re the presence of polymers doesn't affect the nature of extreme events in polymer dissipation, which overlap to the Newtonian curve. This is to say the intermittency corrections are the same in Newtonian and polymeric turbulence at large Re, consistently with the collapse of the multifractal spectra presented by~\citep{Marco23}. The differences begin to surface at smaller Re where extreme events become more prominent, especially at large De. This is again consistent with the kurtosis picture in \cref{fig:Kurt}(b) and (c), where the curves steepen at large De and small Re. A crucial difference with \cref{fig:Dissf} is seen at Re$_\lambda$ = 40, where the lognormal plot shows that the dissipation distributions vary non-monotonically with De, whereas Fig.~\ref{fig:Epsf} shows that the dependence is monotonic.

We show the distributions of the polymer dissipation $\epsilon_{\rp}$ in~\cref{fig:Dissp}, which are sub-log-normal, unlike $\epsilon_{\rf}$, thus indicating a significantly less intermittent behaviour. In particular, at small $De$, $\epsilon_{\rp}$ shows more large deviations. These polymers rarely stretch and even when stretched, they quickly relax back to their small, equilibrium lengths. This means any deviations about a small mean are more significant. Thus, the polymer dissipation $\epsilon_{\rp} \propto	\trC$ shows the widest tails at the smallest $De = 1/9$, which are wider than even a log-normal distribution, especially at larger $\Rel$'s in~\cref{fig:Dissp}(a,b). With increasing $De$ polymers tend to stretch more and for longer duration owing to the larger relaxation times. Thus, fluctuations in lengths become less important at large $De$, and consequently, the stored energy is dissipated away in a more uniform and less intermittent manner, with fewer large deviations. This results in the $\log \epsilon_{\rp}$ distributions becoming progressively sub-lognormal. This effect is rather persistent across all $\Rel$, even into the ET regime at $\Rel \approx 40$, as seen from~\cref{fig:Dissp}. 

\section{Conclusions}
\label{sec:conclusions}
In this work, we investigate how various measures that typically characterise classical, Newtonian turbulence behave in polymeric flows as both elasticity and inertia of the flow are varied, {especially focusing on how the large and small $Re$ regimes connects at intermediate Reynolds numbers}. While turbulence statistics indeed depend on both $Re$ and $De$, which quantify fluid inertia and polymer elasticity, respectively, we show that the addition of polymers has contrasting effects on different turbulence statistics.

We begin by showing that the energy spectrum in polymeric flows shows a {non-unique scaling} nature, especially at large $Re$ and moderate $De$, where three different dominant flux contributions---viz. fluid non-linearity, polymeric and the dissipative contributions---lead to three distinct scaling exponents. This {non-unique scaling} behaviour is invariably lost at both very large and very small polymer elasticity, where instead the Newtonian behaviour is recovered similar to the observation by~\citet{Marco23}. With decreasing $\Rel$, the Newtonian and polymeric regimes are progressively lost as the fluid inertia weakens and viscosity becomes more important. A unique, smooth, dissipative, elastic turbulence regime spans a wide range of scales at very small $\Rel$, when polymer elasticity is large. We further shed light on the dynamics of our polymeric flows by estimating the typical life-times of velocity fluctuations,  and the rates of energy transfer to smaller fluid scales and to the polymeric mode. We use scaling arguments to show that the classical fluid energy cascade is slowed and weakened in the presence of polymers. The real space statistics of velocity fluctuations, captured by the structure functions, also admit different scaling exponents that are directly related to the Fourier space scaling of the energy spectra. 

The real space statistics also additionally reveal the intermittent nature of velocity fluctuations in polymeric flows. At very large $\Rel$, the presence of polymers has almost no effect on the kurtosis of fluctuations. However, with decreasing $\Rel$, elastic stresses become increasingly more important than the fluid non-linearity and result in strongly intermittent velocity fluctuations. The relative importance of polymer stresses is made clear by the end-to-end polymer lengths which is given by the trace of the conformation tensor. As expected, polymers stretch more as their elasticity increases and they are able to span a wide range of scales. We further study how polymers stretch in different regions of the flow by decomposing velocity gradients into extensional and rotational components. Extensional regions, quantified by the symmetrised velocity gradients, promote the stretching of the polymers, whereas the local rotation rates, quantified by enstrophy, play a rather indirect role in their stretching. While we indeed find that local rotation rates correlate with large polymer lengths, since velocity gradients intensify vorticity by stretching them in their direction, polymers are shown to also stretch in regions with small rotation rates, thus indicating a lesser correlation between enstrophy and polymer lengths. 

Finally, we relate polymer lengths and velocity gradients to the energy dissipated away by the carrier fluid and the polymers, and compare them to a log-normal distribution. At large $\Rel$, much like the kurtosis of velocity fluctuation, fluctuations of fluid dissipation remains unchanged upon the addition of polymers. However, polymer stresses become important at small $\Rel$ and result in large deviations in fluid dissipation. The effect of polymers is most clear at the smallest $\Rel$ where the fluid non-linearity is inactive and yet the fluid dissipation distribution is very far from log-normal. On the other hand, dissipation by polymers themselves is much less intermittent compared to the fluid dissipation, with tails shrinking even further as polymer elasticity increases. This is because polymers are able to stretch more uniformly and hence dissipate away their stored energy in a less intermittent, more uniform fashion. {While this analysis relates the local strain and rotation rates to the stretching of polymers, it is known that the Lagrangian history of these quantities can also be crucial in determining the behaviour of polymer stresses, as shown in the works of~\citet{Wagner2016,Manish2023}. Future Lagrangian studies on PHIT should better shed light on how closely polymer stresses mirror the local flow topology across such a wide range of $Re$ and $De$ numbers.} 

With this work, we have connected polymeric turbulence at large Reynolds number with elastic turbulence, where inertia effects are vanishing. While the two regimes show substantial differences, as for example clearly indicated by the {non-unique scaling} behaviour of the energy spectra, the transition between the two is rather smooth and continuous in the chosen framewok of homogeneous and isotropic turbulence. Future work should deal with better characterising and explaining the origin of the various identified exponents of the energy spectra, as well as extending the analysis to more complex flows.

\backsection[Acknowledgements]{
The research was supported by the Okinawa Institute of Science and Technology Graduate University (OIST) with subsidy funding to M.E.R. from the Cabinet Office, Government of Japan. M.E.R. also acknowledges funding from the Japan Society for the Promotion of Science (JSPS), grants 24K00810 and 24K17210. The authors acknowledge the computer time provided by the Scientific Computing \& Data Analysis section of the Core Facilities at OIST, and the computational resources provided by the HPCI System (Project IDs: hp250035, hp230018, hp220099, hp210269, and hp210229). R.K.S. thanks Alessandro Chiarini (OIST) for discussions and suggestions. }

\backsection[Declaration of interests]{The authors report no conflict of interest.}


\bibliographystyle{jfm}
\bibliography{references.bib,ref_SM.bib}

\end{document}